\documentclass[12pt]{article}
\usepackage{amsthm}
\usepackage{amsmath}
\usepackage{natbib}
\usepackage[colorlinks,citecolor=blue,urlcolor=blue,filecolor=blue,backref=page]{hyperref}
\usepackage{graphicx}

\usepackage{enumerate}
\usepackage{multirow}
\usepackage{verbatim}
\usepackage[font=footnotesize]{subcaption}
\usepackage{amsfonts}
\usepackage{amssymb}
\usepackage{color,soul}
\usepackage{hyperref}
\usepackage{tikz}
\usetikzlibrary{decorations.pathmorphing} 
\usetikzlibrary{fit}					
\usetikzlibrary{backgrounds}	

\newcommand\Tstrut{\rule{0pt}{2.6ex}}       
\newcommand\Bstrut{\rule[-0.9ex]{0pt}{0pt}} 
\newcommand{\TBstrut}{\Tstrut\Bstrut} 
\newcommand{\blind}{1}

\begin{document}

\def\spacingset#1{\renewcommand{\baselinestretch}%
{#1}\small\normalsize} \spacingset{1}


\if1\blind
{
  \title{\bf Nonparametric Bayes Differential Analysis of  Multigroup DNA Methylation  Data}
   \author{Chiyu Gu
\\
   Bayer Crop Science,
 Chesterfield, Missouri\\
    and \\
    Veerabhadran\ Baladandayuthapani\\
    Department of Biostatistics, University of Michigan\\
    and \\
     Subharup Guha\\
    Department of Biostatistics, University of Florida}
  \maketitle
} \fi

\if0\blind
{
  \bigskip
  \bigskip
  \bigskip
  \begin{center}
    {\LARGE\bf Causally Interpretable Meta-Analysis of Observational Studies}
\end{center}
  \medskip
} \fi

\bigskip
\begin{abstract}
DNA methylation datasets in cancer studies are comprised of  measurements on a large number of genomic locations  called cytosine-phosphate-guanine (CpG) sites  with complex  correlation structures. 
A fundamental goal of these  studies  is the development of  statistical techniques that can identify  disease genomic signatures   across multiple patient groups defined by different experimental or biological conditions. 
We propose \textit{BayesDiff},  a   nonparametric Bayesian approach for differential analysis relying on a novel class of first order mixture models called the Sticky Pitman-Yor process or  two-restaurant two-cuisine  franchise (2R2CF).  The BayesDiff methodology flexibly utilizes information from all   CpG sites or probes,   adaptively accommodates any serial dependence due to the widely varying inter-probe distances, and performs simultaneous inferences about the differential  genomic signature of the patient groups. Using simulation studies, we demonstrate the effectiveness of  the BayesDiff procedure relative to existing statistical techniques for differential DNA methylation.  The methodology is applied to analyze a   gastrointestinal (GI) cancer   dataset that displays both serial correlations and  interaction patterns. The results  support and complement known aspects of DNA methylation and gene association in upper GI cancers.
\end{abstract}

\noindent%
\vfill

\spacingset{1.2} 

\section{Introduction}\label{S:intro}
Recent advances in array-based and next-generation sequencing (NGS) technologies have revolutionized biomedical research, especially in cancer. The rapid decline in the cost of genome technologies has facilitated the availability of 
 datasets  involving intrinsically different sizes and scales of high-throughput data and  provided genome-wide, high resolution information  about the biology of cancer.
A common analytical goal is  the identification of differential genomic signatures between groups of samples  corresponding to different treatments or biological conditions, e.g., treatment arms, response to adjuvant chemotherapy, tumor subtypes, or cancer stages. The  challenges include the high dimensionality of genomic biomarkers or probes, usually  in the hundreds of thousands, and the relatively small number of patient samples, usually no more than a few~hundred. This ``small $n$, large $p$'' setting results in unstable inferences due to collinearity. Further, there  exist complex interaction patterns, such as signaling or functional pathway-based interactions, and  location-based serial correlation for high-throughput sequencing data. These data attributes  significantly affect  the reliability of statistical techniques for detecting   differential genomic signatures.

\paragraph{Differential DNA methylation in  cancer studies}  
DNA methylation is an  important epigenetic mechanism that involves the addition of a methyl (CH$_3$) group to DNA, resulting in the  modification of  gene functions. It typically occurs  at specific genomic locations called cytosine-phosphate-guanine (CpG) sites. 
Alterations in DNA methylation, e.g., hypomethylation of oncogenes and hypermethylation of tumor suppressor genes, are often associated with the development and progression of cancer \citep{feinberg2004history}.  
It was previously believed that these alterations  occur almost exclusively at   promoter regions known as CpG islands, i.e., chromosomal regions with high concentrations of CpG sites. However, with the advent of high-throughput technologies, it has been shown that a significant proportion of cancer-related  alterations do not occur in    promoters or CpG islands \citep{irizarry2009genome}, prompting higher resolution, epigenome-wide investigations.

Gastrointestinal (GI)  cancer, the most common form of cancer in the U.S. \citep{CAAC:CAAC21387}, refers to malignant conditions affecting the digestive system   associated  
with epigenetic alterations  \citep{vedeld2017epigenetic}. 
Molecular characterization of  different cancer types, facilitated by the  identification of differentially methylated CpG sites, 
 is therefore key to  better understanding   GI cancer.
%
In the motivating application, we  analyze  methylation profiles  publicly available from The Cancer Genome Atlas (TCGA) project, consisting of 1,224   tumor samples belonging to four GI cancers of the upper digestive tract: stomach adenocarcinoma (STAD), liver hepatocellular carcinoma (LIHC), esophageal carcinoma (ESCA) and pancreatic adenocarcinoma (PAAD). For 485,577 probes, where each probe is mapped to a CpG site, DNA methylation levels {or Beta-values} ranging from 0 (no methylation) to 1 (full methylation)  were measured using the Illumina Human Methylation 450 platform. 

\begin{figure}[h]
	\centering
\includegraphics[width=\textwidth]{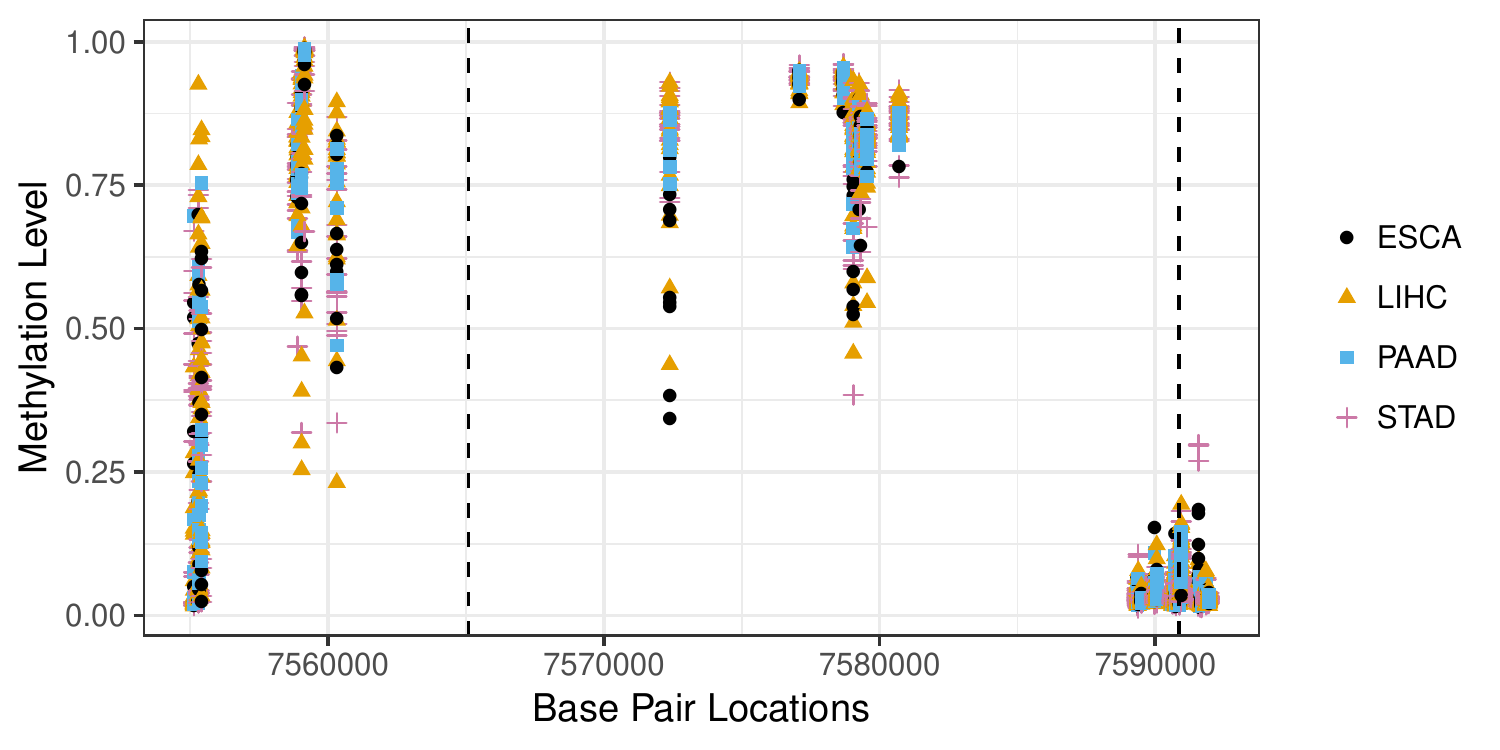}
\caption{Methylation levels of  CpG sites near gene TP53 {for  randomly chosen tumor samples  of the TCGA upper GI dataset}. Each {plotted point represents the methylation level at a CpG site, with the  shapes and colors corresponding to different GI cancers indicated in the legend. The  vertical dashed lines demarcate the  TP53 gene boundaries. }}
	\label{intro_data_plots}
\end{figure}

Figure \ref{intro_data_plots} displays the methylation levels for CpG sites near  TP53, a tumor suppressor gene located on chromosome $17$. A random subset of the tumor samples was chosen to facilitate an informal visual evaluation. Each plotted point represents the methylation level of a  tumor sample at a CpG site. As indicated in the figure legend, the four sets of colors and shapes of the points represent  the four upper GI cancers. The vertical dashed lines indicate the boundaries of the TP53 gene. Although  differential methylation is clearly visible  at some CpG sites, the differences 
are generally subtle, demonstrating the need for  sophisticated statistical analyses.
An obvious  feature  is the correlation between 
the apparent methylation statuses of nearby CpG sites \citep{eckhardt2006dna, irizarry2008comprehensive, leek2010tackling}. The  dependence  of proximal CpG sites is also seen in Figure \ref{intro_data_plots2} of Supplementary Material, where we find   high first order autocorrelations and  highly significant  tests for serial correlations. Furthermore, the  variability of the inter-probe spacings in  Figure~\ref{intro_data_plots}  suggests that the need to model distance-based dependencies.

\paragraph{Existing statistical approaches for differential DNA methylation and limitations}  
Numerous frequentist and Bayesian methods have been developed for  differential DNA methylation,
and  can be broadly classified into four categories: \textit{(i)}
\textit{Testing-based methods}, such as Illumina Methylation Analyzer (IMA) \citep{wang2012ima}, City of Hope CpG Island Analysis Pipeline (COHCAP) \citep{warden2013cohcap}, and BSmooth \citep{hansen2012bsmooth}. These methods rely on two-sample or multiple-sample tests for the    mean group differences at each CpG site. 
\textit{(ii)}
\textit{Regression based models}, such as Methylkit \citep{akalin2012methylkit}, bump hunting \citep{jaffe2012bump}, Biseq \citep{hebestreit2013detection}, and RADMeth \citep{dolzhenko2014using}. After applying smoothing or other adjustments, these methods fit individual regression models for each CpG site and test for  significance. 
\textit{(iii)}
\textit{Beta-binomial model-based methods}, such as MOABS \citep{sun2014moabs}, DSS \citep{feng2014bayesian}, and methylSig \citep{park2014methylsig}. These methods fit separate models to each CpG site.
\textit{(iv)}
\textit{Hidden Markov models (HMMs)}, such as MethPipe \citep{song2013reference}, Bisulfighter \citep{saito2014bisulfighter}, and HMM-DM \citep{yu2016hmm}. These methods   detect differentially methylated sites based on  inferred hidden~states.

The aforementioned  methods have several    deficiencies. By fitting
separate models to each probe, most   methods ignore the strong correlations between neighboring probes. This   
 reduces the  detection power 
for the relatively small sample sizes. Additionally, beta-binomial, HMM, and most  testing-based methods are able to accommodate only two treatments and  rely on inefficient 
 adjustments to compare multiple treatments. 
 The  methods that  account for serial dependence (e.g., HMMs) do not adjust for the widely varying  distances between the  probes, and   instead, assume  uniform  inter-probe dependencies. 
The few methods that  account for inter-probe distances \citep[e.g.,][]{hansen2012bsmooth, jaffe2012bump, hebestreit2013detection}  rely on ad hoc parameter-tuning procedures that do not  adjust for  distinctive data characteristics. 


Motivated by these challenges, we propose general and flexible methodology for differential analysis in DNA methylation data,
 referred to  as \textit{BayesDiff}.
Rather than fitting a separate model  to   each CpG site or probe, BayesDiff relies on a global analytical framework for   simultaneous inferences  on  the probes that adapts to the unique data attributes. 
To diminish  collinearity effects and achieve dimension reduction,  the probes are allocated to a smaller, unknown number of
latent clusters based on the  similarities of probes-specific  multivariate  parameters. Finally, differential state  variables of the probes delineate the genomic signature of the disease to fulfil the main inferential goal. 

For realistically modeling the  probe-cluster allocation mechanism of DNA methylation profiles,
 we devise an extension of  Pitman-Yor processes (PYPs) \citep{perman1992size} called the \textit{Sticky PYP} (equivalently, the \textit{two-restaurant two-cuisine  franchise}). In addition to accounting for long-range biological interactions, this  nonparametric  process  accommodates distance-based serial dependencies of the probes. 
 Separately for the differential and non-differential probes, it flexibly permits the data to  direct the choice
between PYPs, and their special case, Dirichlet processes, in  finding the best-fitting allocation schemes. 

We implement an  inferential procedure for Sticky PYPs  using a Markov chain Monte Carlo (MCMC) algorithm  specifically designed for posterior inferences in the typically  large methylation datasets. 
Simulation results show that our approach significantly outperforms existing methods for multigroup comparisons in  data with or without serial correlation. For the motivating TCGA  dataset, in addition to confirming known features of DNA methylation and disease-gene associations, the analysis reveals  interesting aspects of  the biological mechanisms of upper GI~cancers.

The rest of the paper is organized as follows. Section \ref{model_desc_section} describes the BayesDiff approach, with  Section~\ref{Sticky_PYPs_DA} introducing the Sticky PYP or two-restaurant two-cuisine  franchise (2R2CF) for differential DNA methylation. 
Section~\ref{S:post_inf} outlines an  effective  inference procedure for  detecting  differential probes.
  Section \ref{sim_study} uses artificial datasets with varying noise and correlation levels to assess the accuracy of BayesDiff in detecting  disease genomic signatures and makes comparisons  with established techniques for DNA methylation data.
The motivating upper GI dataset is analyzed using the BayesDiff procedure in Section \ref{data_analysis}.  Finally, conclusions and future related work are discussed in Section~\ref{disc}.

\section{The BayesDiff Model} \label{model_desc_section}

Sequencing technologies measure
DNA methylation levels of $p$ biomarkers represented by CpG sites (``probes'') and $n$ matched
patient or tissue samples (``individuals''). Usually,   $p$ is  much larger than $n$. The  methylation levels, which belong to the interval $[0,1]$,  are  arranged in an $n \times p$ matrix of  proportions, $\mathbf{X}=((x_{ij}))$, for individuals $i$ and probes $j$, with the probes  sequentially indexed by their genomic locations.
The distances between  adjacent probes 
are denoted by $e_1,\ldots,e_{p-1}$, and typically  exhibit high variability. For instance, 
in the upper GI TCGA dataset,  the  inter-probe distances range from 2 base pairs to  a million base pairs;  a base pair is a  unit of DNA length consisting of two nucleobases bound to each other by hydrogen bonds \citep[e.g.,][]{baker2008molecular}.

Each individual $i$ is associated with a known experimental or biological condition (``treatment'') denoted by $t_i$ and taking values in $\{1,...,T\}$ with  $T\ge 2$. In the motivating TCGA data, there are $T=4$ upper  GI cancer~types. 
We model  the  logit
 transformation of the methylation levels, $z_{ij}=\log\left(x_{ij}/\left(1-x_{ij}\right)\right)$, as follows:  
\begin{equation}
z_{ij} \sim N\left(\xi_i+\chi_j+\theta_{t_ij},\sigma^{2}\right)
\label{def_main_model}
\end{equation}
where $\xi_i$ represents the $i$th subject's random effect, $\chi_j$ represents the $j$th probe's random effect, and  $\theta_{tj}$ 
is the random treatment $t$--probe $j$ interaction effect. The logit-transformed methylation levels differ from   M-values \citep{du2010comparison}, commonly used in differential analyses,     by  $1-\log(2)=0.306$; however, the key results are identical.

The main inferential  goal is the detection of differential probes, i.e.,   probes $j$ for which the elements of column vector $\boldsymbol{\theta}_j=$ $\left(\theta_{1j},\dots, \theta_{Tj}\right)'$ are not all identical. 
  Consequently, we define a binary \textit{differential state variable}, $s_j$, with $s_j=1$  indicating that probe $j$ is not  differential and $s_j=2$ indicating that it is differential:
\begin{align}
s_j = 
\begin{cases}
1  \qquad\text{if } \theta_{1j}=\theta_{2j}=\dots=\theta_{Tj}, \\
2 \qquad \text{otherwise},
\end{cases}
\label{def_nonDE}
\end{align}
for  $j=1, \dots, p$. Thus, the key parameters are   $s_1,\ldots,s_p$, with
the differential genomic signature consisting of the probes  for which state $s_j=2$.  Motivated by the distance-dependent correlations of DNA methylation data and  deficiencies of  existing statistical approaches,   this paper fosters a  Bayesian nonparametric framework for   random effects $\boldsymbol{\theta}_1,\ldots,\boldsymbol{\theta}_p$  determining the differential state variables.

\subsection*{Modeling probe  clusters} \label{PYP_sec_main}
As mentioned, in addition to high-dimensionality, the analytical challenges include pervasive collinearity due to   dependencies between physically proximal probes. Additionally, there are long-range dependencies between non-adjacent probes because of  biological interactions, e.g., signaling or functional pathways. 
 To accommodate  these dependence structures  and  extract  information from the large number of probes, we allocate the $p$ probes to a much smaller number, $q$, of latent clusters based on  the  similarities  of the random effects $\boldsymbol{\theta}_j$. We favor  clustering  to    dimension reduction methods such as principal components analysis (PCA). Because each principal component is a linear combination  of all $p$ biomarkers, PCA  is less useful  in cancer research because of its inability to select features, i.e., probes. 
By contrast,   our approach facilities  biological interpretations by  identifying  CpG sites   relevant to the differential  genomic~signatures between the treatments.

Suppose that an \textit{allocation variable}, $c_j$, assigns probe $j$ to one of  $q$ latent clusters, where $q$ is unknown. The event $[c_j=k]$ signifies that the $j^{th}$ probe belongs to the $k^{th}$ latent cluster,  $k=1,\dots, q$.
We assume that the $q$ clusters are  associated with  \textit{latent vectors}, $\boldsymbol{\lambda}_1,\dots, \boldsymbol{\lambda}_q$, where the probe-specific random effects and cluster-specific latent vectors have the relation:
\begin{equation}
\boldsymbol{\theta}_j=\boldsymbol{\lambda}_k \quad \text{if $c_j=k$.}\label{allocation c} 
\end{equation}
That is, all  probes within a cluster are assumed to have identical random effects equal to that cluster's latent vector. The differential state variables of the probes, defined in equation (\ref{def_nonDE}), then become a shared  attribute of their parent   cluster, and  clusters as a whole are either differentially or non-differentially methylated. Further,  if probe $j$ belongs to cluster $k$ (i.e., $c_j=k$), then the  condition $\theta_{1j}=\theta_{2j}=\dots=\theta_{Tj}$ in equation~(\ref{def_nonDE}) is   equivalent to  $\lambda_{1k}=\lambda_{2k}=\dots=\lambda_{Tk}$, and 
  differential clusters constitute the set
\begin{equation}
    \mathcal{D} = \left\{k: \lambda_{tk} \neq \lambda_{t^{'}k}, \text{ for some } t \neq t^{'}, k=1,\ldots,q\right\}. \label{D}
\end{equation}

\paragraph{Mixture models for  allocation}  \quad 
 Bayesian infinite mixture  models are a natural choice  for allocating  $p$ probes to a smaller, unknown number of latent clusters based on their random effects similarities. 
Dirichlet processes \citep{ferguson1973bayesian} are arguably the most frequently used infinite mixture models; see \citet[chap.\ 4]{muller2013bayesian} for a comprehensive review. The use of Dirichlet processes  to achieve
 dimension reduction  has precedence in the literature, albeit in unrelated applications \citep[see][]{Medvedovic_etal_2004,Kim_etal_2006, Dunson_etal_2008,Dunson_Park_2008,Guha_Baladandayuthapani_2016}.
 \cite*{Lijoi_Mena_Prunster_2007b}  advocated the use of Gibbs-type priors 
 \citep*{gnedin2006exchangeable} for accommodating more flexible clustering mechanisms and    demonstrated the  utility of Pitman-Yor processes (PYPs) in genomic applications. 
 An overview of Gibbs-type priors and  characterization of the learning mechanism is provided by \cite{de2013gibbs}. Formally, the PYP \citep{perman1992size} relies on a discount parameter $d \in [0,1)$,  positive mass parameter $\alpha$, and  $T$-variate base distribution $W$, and is denoted by $\mathcal{W}(d,\alpha,W)$. 
The value $d=0$ yields a Dirichlet process with mass parameter $\alpha$ and  base distribution $W$. Suppose  $\boldsymbol{\theta}_1, \ldots, \boldsymbol{\theta}_p$ are distributed as $\mathcal{W}(d,\alpha,W)$. The \textit{stick-breaking representation} of  $\mathcal{W}(d,\alpha,W)$ is  $\boldsymbol{\theta}_j \stackrel{iid}\sim \mathcal{P}$, where  random  distribution $\mathcal{P}$ is the discrete mixture $\sum_{v=1}^\infty \pi_v \delta_{\boldsymbol{\phi}_v}$, with $\delta_{\boldsymbol{\phi}_v}$ denoting a point mass located at the atom $\boldsymbol{\phi}_v \stackrel{iid}\sim W$. The random stick-breaking probabilities have the form $\pi_1 = V_1$, and $\pi_h = V_h \prod_{v=1}^{h-1}(1-V_v)$ for $h > 1$, where $V_h \stackrel{indep}\sim \text{beta}(1-d, \alpha+hd)$.  \cite{Guha_Baladandayuthapani_2016} introduced VariScan, a technique that utilizes  PYPs and Dirichlet processes for  clustering, variable selection, and prediction in high-dimensional regression problems in general, and in gene expression datasets in particular. They also demonstrated that PYPs are overwhelmingly favored to Dirichlet processes in gene expression  datasets, which typically exhibit no  serial~correlation.

\paragraph{Limitations of existing mixture models}  \quad Although the aforementioned  mixture models achieve dimension reduction  and account for the long-range biological interactions between non-adjacent probes, 
a potential drawback  is their implicit assumption of a priori  exchangeability of the probes. Consequently, these techniques cannot account for the serial correlation  in  methylation data. 
Infinite HMMs, such as the hierarchical Dirichlet process hidden Markov model (HDP-HMM)  \citep{Teh_etal_2006} and  Sticky HDP-HMM \citep{Fox_etal_2011}, may be utilized to fill this gap. 
 Although these models are a step in the right direction,   they have several undesirable features for differential analysis. \textit{First}, the degree of first order dependence  is uniform irrespective of the inter-probe distances. This is  unrealistic in methylation datasets where the  correlation typically decreases with inter-probe distance \citep{hansen2012bsmooth, jaffe2012bump, hebestreit2013detection}.
{\textit{Second}, 
an ad hoc exploratory analysis of the  GI cancer dataset  reveals that the serial correlation in the treatment-probe effects  is    weaker than the serial dependence between  the  differential state variables defined in equation (\ref{def_nonDE}). 
Although there may not  be a  biological explanation for this phenomenon, this makes 
 sense from a statistical perspective because  the differential states are binary functions of the  treatment-probe interactions; the  differential states  are  more sensitive in detecting first order dependence even when the higher-dimensional (and noisier) treatment-probe interactions show little or no correlation.} 
This suggests that a hypothetical two-group Markov model, rather than  an infinite-group Markov model such as HDP-HMM or Sticky HDP-HMM, would  provide a  better fit for the data. 
\textit{Third}, the range of allocation patterns supported by infinite HMMs is relatively limited. In particular, realistic allocation patterns, such as power law decays in the cluster sizes and  large numbers of small-sized clusters, a common feature of cancer datasets \citep{Lijoi_Mena_Prunster_2007a}, are assigned relatively small prior probabilities by infinite HMMs.


\subsection{Sticky PYP: A Two-restaurant, Two-cuisine Franchise (2R2CF)  for Differential Analysis}\label{Sticky_PYPs_DA} 

We invent a  mixture model called the Sticky PYP. Essentially, this is   a cohort of regular PYPs that generates the probe-specific random effects by switching  generative PYPs at random locations along the probe sequence. Alternatively, the well-known Chinese restaurant franchise (CRF) metaphor for HDP-HMMs and Sticky HDP-HMMs  can be generalized to the \textit{two-restaurant two-cuisine  franchise} (2R2CF) to  provide an equivalent representation of  Sticky PYPs appropriate for differential analysis. We  first present a descriptive overview of 2R2CF.

Imagine that a franchise has two restaurants, labeled  1 and 2. Each restaurant consists of two sections, also labeled  1 and 2. Each section  serves a single cuisine and its  section-cuisine menu  consists of  infinite dishes. Section 1   at both   restaurants exclusively serves  cuisine~1. The cuisine~1 menus, along with  selection probabilities associated with the dishes,  are identical at  the  two restaurants. Similarly,  section 2 at both restaurants  exclusively serves  cuisine~2, and the cuisine 2 menus  are also identical at the restaurants. 

A succession of $p$ customers, representing the  CpG sites or probes, arrive at the franchise. The waiting times between successive customers represent the inter-probe distances, $e_1, \dots,e_{p-1}$.  
Each customer selects a restaurant and then selects a section (equivalently, cuisine) in the restaurant.  
 Each  restaurant section has an infinite number of tables, and a  customer   may either sit at a table already occupied by previous customers, or  sit at a new table. 
 All customers  at a  table are  served the same dish, previously  chosen from the section-cuisine menu  by the first customer who sat at that table. A customer who sits at a new table is allowed to independently pick a dish from the cuisine menu with a cuisine-specific probability associated with each dish. As a result, multiple tables in a restaurant section may serve the same dish. 
 
 Restaurant~1 specializes in  cuisine~1.  Consequently,  section 1 is  more popular with  restaurant~1  patrons. Similarly, restaurant 2 specializes in cuisine 2, and so,    restaurant~2 customers tend to favor section 2 over section 1. 
  By design, if a  customer has eaten a  cuisine 1 (2) dish, then the next customer is more likely to visit restaurant 1 (2), where cuisine 1 (2) is more popular. In this manner, each customer tends to select the same cuisine as the previous customer.
 
 In the metaphor, cuisine 1 symbolizes the  {non-differential  state}  and cuisine 2 symbolizes the  {differential  state}. The dish that  franchise customer $j$ eats  represents  the  probe-specific random effect, $\boldsymbol{\theta}_j$. 
Since cuisine 1 represents the  {non-differential  state},  its  dishes are characterized by $T$-variate random  vectors with all equal elements; see equation~(\ref{def_nonDE}). In contrast, cuisine 2 (differential  state) dishes  are characterized by $T$-variate random vectors with at least two unequal elements.  

The dependence in the restaurant and cuisine choices of consecutive customers  account for  long runs  of differential or  non-differential states in DNA methylation data. 
However, a customer's influence on the next customer  diminishes  as the  time interval between    the two customers~increases. That is, the differential statuses of  two adjacent probes become  statistically independent in the limit as the inter-probe distance~grows.

The 2R2CF process  is  illustrated in Figure~\ref{2R2CF plots} and discussed below in greater~detail. 
The following specification conditions on $G$, an unknown distribution in $\mathcal{R}$ that is assigned  a  Dirichlet process  prior  with  mass parameter $\beta>0$ and univariate normal base distribution, $G_0=N\left(\mu_G,\tau_G^{2}\right)$.
 The stick-breaking representation of the Dirichlet process implies that  distribution $G$ is almost surely discrete because it has an infinite mixture distribution:
\begin{equation}
G \stackrel{d}= \sum_{v=1}^\infty \varpi_{v} \delta_{\zeta_{v}}, \text{ where $\sum_{v=1}^\infty \varpi_{v}=1$ and $\zeta_{v} \stackrel{iid}\sim
G_0$}. \label{density G}
\end{equation}
The  distribution of  random  probabilities,  $\varpi_{v}$, which depend on mass parameter $\beta$, was derived in \cite{sethuraman1994constructive}; see also \cite{Ishwaran_James_2003} and \cite{Lijoi_Prunster_2010}. In the sequel, we condition on  distribution $G$; equivalently on the probabilities,  $\varpi_{v}$, and univariate atoms,  $\zeta_{v}$, for  $v\in\mathcal{N}$.


\paragraph{Cuisine 1 menu.}  \quad Recall that cuisine 1 represents the non-differential state, for which the  $T$-variate random vectors (i.e., dishes in the metaphor) has all equal elements. Cuisine 1 menu, with its countably infinite dishes and their associated probabilities, is modeled as follows as a discrete $T$-variate  \textit{menu distribution}, $W_1$. With $\boldsymbol{1}_T$ denoting the column vector of $T$ ones, let
\begin{equation}
   \boldsymbol{\vartheta} \mid \psi  = \psi\boldsymbol{1}_T  \text{ where $\psi  \sim G$.}\label{W1}
   \end{equation}
  Cuisine 1 menu distribution, $W_1$, is defined as the law of random vector $\boldsymbol{\vartheta}$. 
 Then $\mathcal{S}_1=$ $\{\zeta_{v}\boldsymbol{1}_T : v \in \mathcal{N}\}$  is the list of available cuisine 1 dishes and  the support of $W_1$. The selection probability associated with  dish $\zeta_{v}\boldsymbol{1}_T$ is $\varpi_{v}$. 
 
 The continuity of base distribution $G_0$ in equation (\ref{density G})  guarantees that   the menu $W_1$ dishes are unique.
The discreteness of distribution $G$  has practical implications for 2R2CF:   \textit{(a)} cuisine 1   consists of  discrete dishes, as required, rather than a continuous spectrum, and \textit{(b)} since section 1 at both the restaurants serve the same menu, the section~1 customers may eat the same dish even if they  select different restaurants.


\paragraph{Cuisine 2 menu.}  As mentioned earlier, cuisine 2 depicts the  {differential  state} and  its dishes  represent $T$-variate random vectors with at least two unequal elements. Its  menu comprises countably infinite cuisine 2 dishes along with their associated probabilities. The menu  is therefore modeled by a $T$-variate    distribution, $W_2$,  satisfying two  conditions: \textit{(i)}~it has a countably infinite support, and   \textit{(ii)} each $T$-variate atom of $W_2$ has at least two  unequal elements.
   For any given $\boldsymbol{\phi}=(\phi_1,\ldots,\phi_T)' \in \mathcal{R}^T$, a  probability mass function for $W_2$ that satisfies  these  conditions is 
   \begin{align}
W_2(\boldsymbol{\phi}) =
\begin{cases}
\prod_{t=1}^T G(\phi_t)/\bigl(1-\sum_{v=1}^\infty \varpi_{v}^T\bigr)  &\qquad\text{if $\phi_{t}\neq \phi_{t'}$ for some $t\neq t'$,}\\
0 &\qquad\text{otherwise,}
\end{cases}\label{density W2}
\end{align}
where $G(\phi)$ denotes the  mass function of distribution $G$, represented in equation~(\ref{density G}),  evaluated at $\phi\in \mathcal{R}$. {In line 1 of expression (\ref{density W2}), the normalizing constant $(1-\sum_{v=1}^\infty \varpi_{v}^T)$ is  the total probability that a $T$-variate random vector  whose elements are i.i.d.\ $G$ has at least two distinct  elements.} Then menu 
distribution $W_2$  satisfies conditions \textit{(i)} and \textit{(ii)}, as required. 
Referring back to the atoms, $\zeta_{v}$, of distribution $G$ in (\ref{density G}),  the list of  cuisine 2 dishes is  
$\mathcal{S}_2= \{\boldsymbol{\phi} : \phi_t=  \zeta_{v} \text{ for some } v \in \mathcal{N}, \text{ but not all $\phi_t$ are  identical}, t=1,\ldots,T\}
$
and  is the support of $W_2$. The selection probability associated with a dish $\boldsymbol{\phi}$ is~$W_2(\boldsymbol{\phi})$. 


\subsubsection{Restaurant, section, table, and dish choices  of the 2R2CF customers}

Let the  restaurant chosen by franchise customer $j$  be denoted by $g_j$ and the chosen cuisine (i.e.,~section)  be  denoted by $s_j$. Suppose he or she sits at table $v_j$ in that restaurant section and eats dish $\boldsymbol{\theta}_j$.

\paragraph{Customer 1.}  At time $0$, suppose  the first  customer   selects restaurant $g_1=1$ with  probability $\rho_1>0$ and  restaurant $g_1=2$ with positive probability $\rho_2=1-\rho_1$. For reasons that  will become clear, we refer to   $\rho_1$ as the \textit{baseline non-differential proportion} and $\rho_2$ as  the  \textit{baseline differential proportion}.  Typically, the differential state is relatively less frequent, and so  $\rho_2<\rho_1$ (i.e.,~$\rho_2<1/2$). Proportion $\rho_1$ is given a  uniform prior on the interval $(0.5,1)$.  

 \textbf{\textit{Choice of cuisine $s_1$}} \quad  
Next, customer 1 selects a section in restaurant $g_1$.
As described earlier, each restaurant specializes in its namesake cuisine, which  is therefore more popular with its customers. This  is modeled as follows.  Within    restaurant $g_j$ (where $j=1$ for the first customer),   customer $j$ selects cuisine 1 with probability 
\begin{align}
\mathcal{Q}_{g_j}(1) =
\begin{cases}
\rho_1 + \rho_2 \gamma  &\qquad \text{if $g_j=1$,}\\
\rho_1 - \rho_1 \gamma  &\qquad \text{if $g_j=2$,}\\
\end{cases}
\label{PYP_state_prob1}
\end{align}
for a \textit{speciality cuisine popularity parameter}, $\gamma \in (0,1)$, that determines the  degree to which a restaurant's patrons favor its  namesake cuisine. For instance,  if  $\gamma$ is nearly equal to 1, then a   restaurant 1~(2) customer almost always (never) chooses  cuisine~1. At the other extreme, if  $\gamma$ is nearly equal to 0, then the customer always chooses  cuisine 1 with approximate probability $\rho_1$ irrespective of the restaurant. Parameter $\gamma$ is assigned an independent uniform prior on the unit interval. The probability that a restaurant~$g_j$ customer chooses cuisine 2 is then $\mathcal{Q}_{g_j}(2)=1-\mathcal{Q}_{g_j}(1)$. 

\textbf{\textit{Choice of table $v_1$ and dish  $\boldsymbol{\theta}_{1}$}} \quad  
 Within section $s_1$ of restaurant $g_1$, since the table identifiers   are arbitrary, 
  we  assume without loss of generality that customer 1 sits at table $v_1=1$. As the 2R2CF process evolves,  the   tables in a restaurant's section are assigned  consecutive labels as newly  arriving  patrons  occupy them. At table $v_1=1$, customer~$1$ randomly orders a cuisine $s_1$ dish from  menu distribution $W_{s_1}$. The  dish he or she eats represents the random effect  of the first probe. That is,  $\boldsymbol{\theta}_{1}\mid s_1 \sim W_{s_1}$.

\begin{figure}
	\centering
	\vspace{-.5 in}
	
\includegraphics[width=0.75\textwidth]{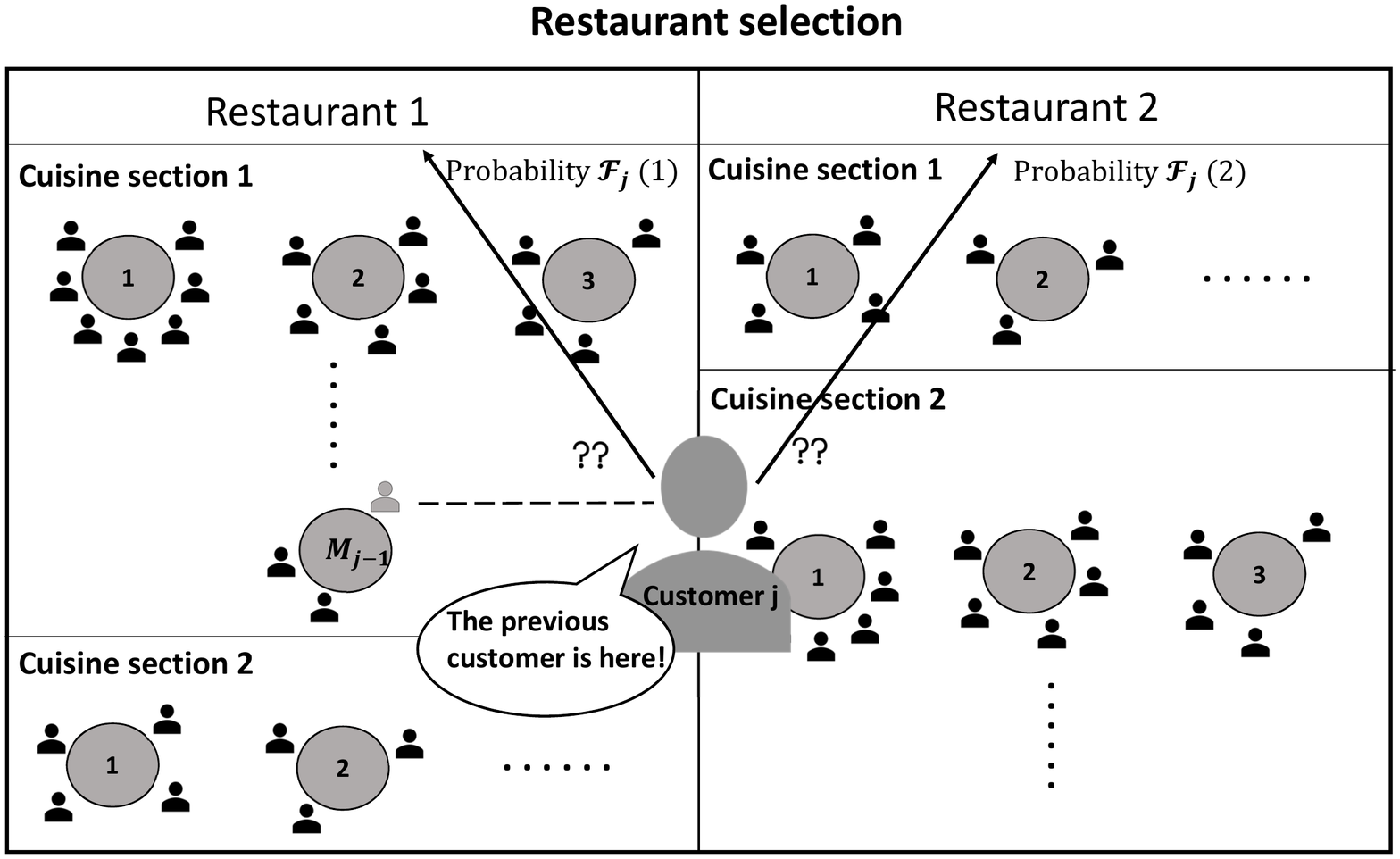}
\vspace{-1 in}

\includegraphics[width=0.75\textwidth]{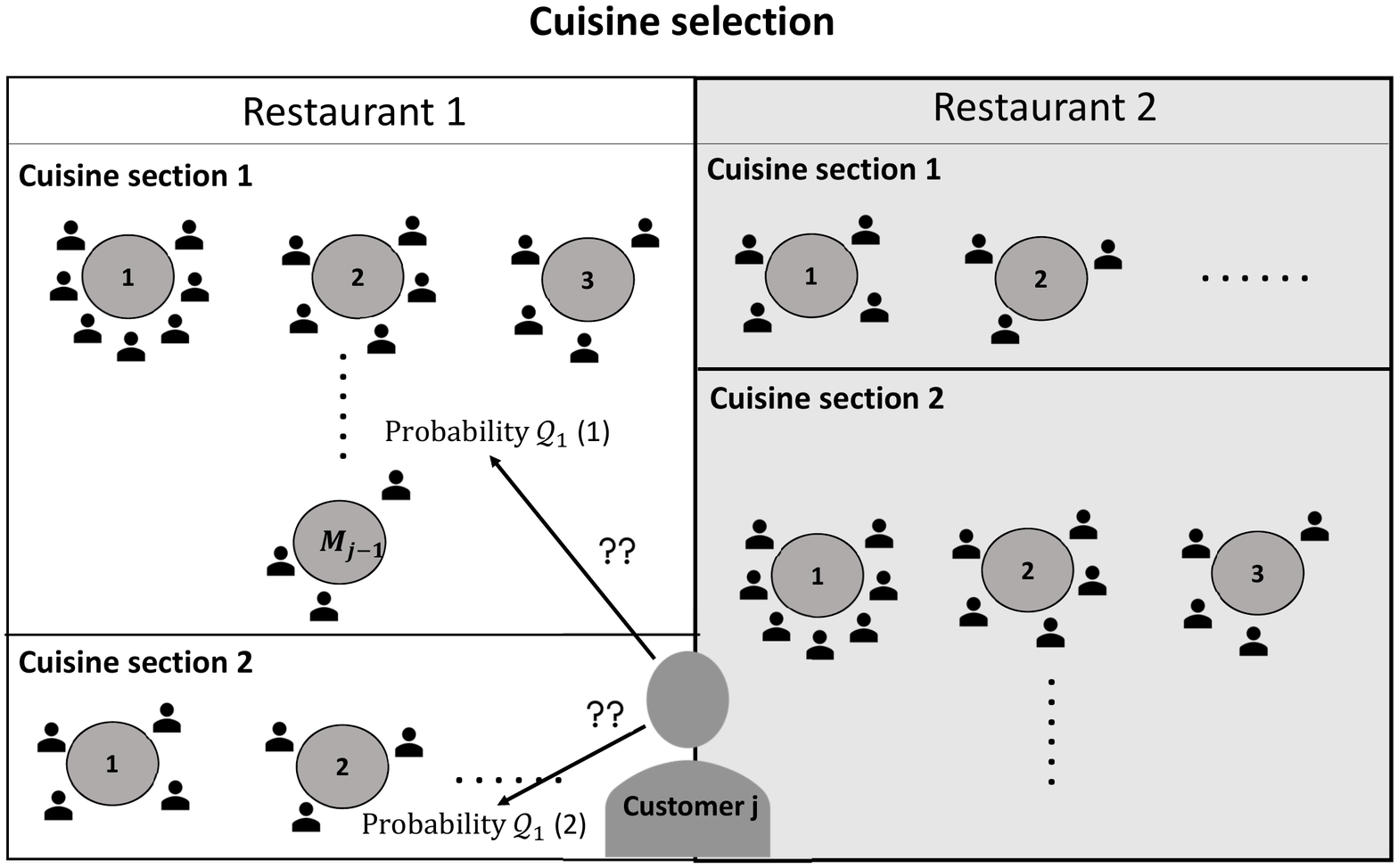}
\vspace{-1 in}

\includegraphics[width=0.75\textwidth]{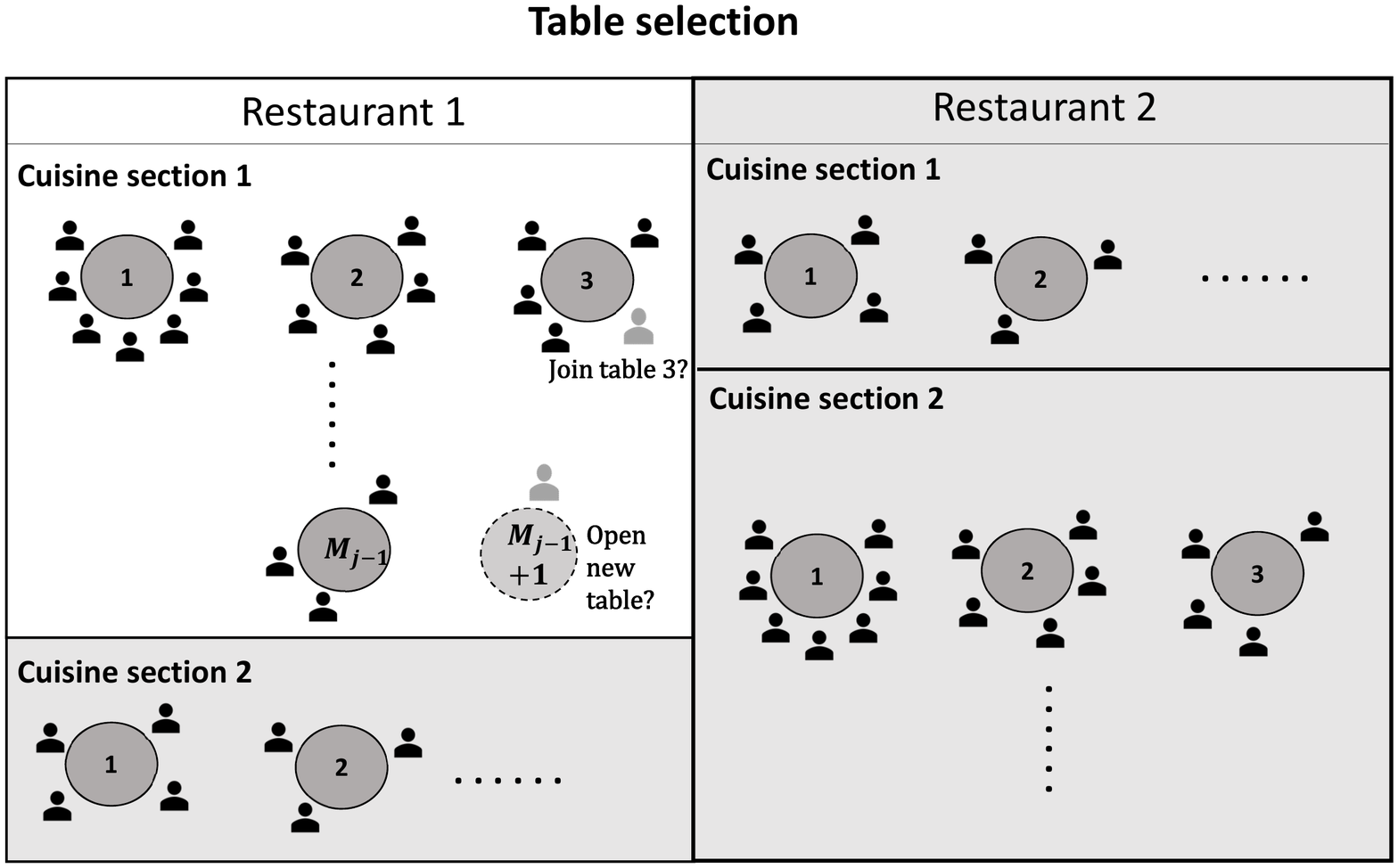}
\vspace{-.5 in}

	\caption{\small Cartoon representation of the two-restaurant two-cuisine franchise for  differential analysis, showing the progressive choice of restaurant, cuisine section, and table by customer $j$, for $j>1$. The numbered circles represent  table numbers. See the text in Section \ref{Sticky_PYPs_DA}  for a detailed description of the process. }
	\label{2R2CF plots}
\end{figure}

\paragraph{Customer $j$, for $j>1$.}  The \textit{restaurant} choice of a subsequent customer is influenced by the  previous customer's \textit{cuisine} and the waiting time between the  customers. Suppose customer $j$ arrives at the franchise after a  time interval of $e_{j-1}$ following the $(j-1)$th customer. Without loss of generality,   $e_1,\ldots,e_{p-1}$  are scaled so that their total  equals~1. Since the probes in differential analysis typically represent CpG sites  on a chromosome or gene,  it  has   a  scaled  length of 1. 

To model the dependencies between  the franchise customers, we first define 
 a non-negative \textbf{\textit{dependence parameter}} 
 $\eta$ that transforms  the waiting time $e_{j-1}$   to  an \textbf{\textit{affinity}} measure between customer $(j-1)$ and customer $j$:
\begin{align}
r_j=\exp(-e_{j-1}/\eta), \qquad j>1.
\label{def_dep_par}
\end{align}
 The affinity measure belongs to the unit interval when $\eta>0$. If  $\eta=0$,   affinity  $r_j$ is defined as $0$ irrespective of  the  waiting~time. The affinity  influences the behavior of customer $j$ through  equation  (\ref{PYP_grp_prob1})  below.

\textbf{\textit{Choice of restaurant $g_j$}} \quad The cuisine $s_{j-1}$ of the $(j-1)$th customer influences the restaurant  choice of the $j$th customer through  affinity  $r_j$ and popularity parameter~$\gamma$: 
 \[g_j \,\big|\, s_{j-1}, \rho_1, \eta, \gamma\sim \mathcal{F}_j.
 \]
 Specifically, the  probability that customer $j$  selects  restaurant 1 is  assumed to be 
   \begin{align}
   \mathcal{F}_j(1) \stackrel{def}= 
P\bigl(g_j = 1 \mid s_{j-1}, \rho_1, \eta, \gamma\bigr) =
\begin{cases}
\rho_1 + \rho_2 r_j /\gamma &\qquad\text{if  $s_{j-1}=1$,}\\
\rho_1 - \rho_1 r_j /\gamma  &\qquad\text{if  $s_{j-1}=2$,}\\
\end{cases}
\label{PYP_grp_prob1}
\end{align}
and $\mathcal{F}_j(2)=1-\mathcal{F}_j(1)$.  
If dependence parameter $\eta=0$, we find that the restaurant choices of the customers are independent; specifically,  $\mathcal{F}_j(1) =\rho_1$   irrespective of the cuisine $s_{j-1}$. The  idea is illustrated in the top panel of Figure \ref{2R2CF plots}, where customer $j$ chooses restaurant 1 with probability $\mathcal{F}_j(1)$ and  restaurant 2 with probability $\mathcal{F}_j(2)$. 

It can be verified that $\mathcal{F}_j$  is a  probability mass function if and only if $r_j/\gamma<1$. 
Since the scaled waiting times are bounded above by 1, a globally sufficient  condition is  $\eta < -1/\log \gamma$. We, therefore, assume a mixture prior for dependence parameter $\eta$:
\begin{align}
\eta \mid \gamma \sim \frac{1}{2}\delta_{0} + \frac{1}{2} \mathcal{H}\cdot \mathcal{I}(\eta<-1/\log \gamma), \label{etaPrior}
\end{align}
where the second mixture component  involves a continuous distribution, $\mathcal{H}$, restricted to the interval $[0,-1/\log \gamma)$, thereby enforcing the  globally sufficient condition.  In our experience, posterior inferences on $\eta$ are relatively robust to the continuous prior  $\mathcal{H}$ provided the prior is not highly concentrated on a small part  of  interval $[0,-1/\log \gamma)$.


When $\eta=0$, we have a \textit{zero-order Sticky PYP};  when $\eta>0$, we obtain a \textit{first order Sticky PYP}.
Some 
 interesting consequences of specification~(\ref{PYP_grp_prob1})  are:

\begin{enumerate}
\item \textbf{Zero-order Sticky PYP}: When $\eta=0$,  each customer independently chooses restaurant 1 (or 2)  with a baseline probability of $\rho_1$ (or $\rho_2$). The $p$ customers  act identically. 

    \item \textbf{First order Sticky PYP  with  $e_{j-1}/\eta$ large}: \quad
        At large relative distances,  customer $j$  acts approximately independently of the  history. Somewhat similarly to customer 1, customer $j$ chooses restaurant 1 (2)  with a  probability approximately, but not exactly, equal to  $\rho_1$ ($\rho_2$). 
    
   \item \textbf{First order Sticky PYP  with  $e_{j-1}/\eta$ small}: \quad
    In the limit as $e_{j-1}/\eta \to 0$ (e.g., for a small inter-probe distance $e_{j-1}$),  the  restaurant  choice of customer~$j$  follows a hidden Markov model.
\end{enumerate}
    
Since it drives the dependence characteristics of DNA methylation data,   parameter  $\eta$ is  of interest. Prior specification~(\ref{etaPrior})  allows the
   data to direct the  model order  through     posterior probability, $P[\eta = 0 \mid \boldsymbol{X}]$, and an MCMC  probability  is readily available.

\textbf{\textit{Choice of cuisine $s_j$}} \quad 
Upon entering    restaurant $g_j$, customer $j$ selects cuisine-section $s_j$ with distribution $\mathcal{Q}_{g_j}$, defined previously  in expression  (\ref{PYP_state_prob1}). For bookkeeping reasons, among  franchise customers $1,\ldots,j$, suppose   $p_{gs}^{(j)}$ customers  choose  section $s$ in restaurant $g$; that is,   $p_{gs}^{(j)}=$  $\sum_{l=1}^{j}\mathcal{I}(g_l=g, s_l=s)$ for $g,s=1,2$.

For a graphical depiction of  cuisine selection by the $j$th customer, see the middle panel of Figure \ref{2R2CF plots}, where $g_j=1$. That is, customer~$j$,  having already chosen restaurant $1$,    now chooses a cuisine-section. Restaurant 2 has been greyed out  because it is no longer accessible to this customer. In the lower panel of Figure \ref{2R2CF plots}, we find that the customer  picked cuisine-section 1, so that $s_j=1$. 

\textbf{\textit{Choice of   table $v_j$}} \quad Applying the above notation,
among the previous  customers $1,\ldots,(j-1)$,  there are  $p_{g_js_j}^{(j-1)}$ customers in the same restaurant and section as the $j$th customer. Suppose these   customers have occupied   tables $1,\ldots,M_{g_js_j}^{(j-1)}$, and that there are
 $p_{g_js_jk}^{(j-1)}$  customers  seated   at the  $k$th table. Let $\mathcal{M}_{j-1}=\bigl\{p_{g_js_jk}^{(j-1)}:k=1,\ldots,M_{g_js_j}^{(j-1)}\bigr\}$ denote the aggregated table occupancies. 

Recall that the newly arrived $j$th customer  may  sit at any of the $M_{g_js_j}^{(j-1)}$ occupied tables  or   a new $(M_{g_js_j}^{(j-1)}+1)$th table. Two possibilities are illustrated in the  lower panel of Figure \ref{2R2CF plots}. 
For a PYP with mass parameter $\alpha_{s_j}$ and cuisine-specific discount parameter $d_{s_j} \in [0,1)$, the  predictive distribution of table $v_j$ of customer $j$ is assumed to be related to the  table occupancies:
\begin{align}
P\biggl(v_j = k \mid  \mathcal{M}_{j-1}\biggr) \propto
    \begin{cases}
        p_{g_js_jk}^{(j-1)} - d_{s_j} 
         \quad &\text{if $k = 1,\ldots,M_{g_js_j}^{(j-1)}$,}\\
        \alpha_{s_j}  + M_{g_js_j}^{(j-1)}  d_{s_j}  \quad &\text{if $k = (M_{g_js_j}^{(j-1)}  + 1)$,}\\
        \end{cases}\label{pred.table_j}
\end{align}
where the second line corresponds to  customer $j$ sitting at a new table, in which  case the new number of occupied tables is $M_{g_js_j}^{(j)} =M_{g_js_j}^{(j-1)} +1$ and table index  $v_j =M_{g_js_j}^{(j)}$. Otherwise, if customer $j$ sits at a previously occupied table, then their table index $v_j \le M_{g_js_j}^{(j-1)}$ and the number of occupied tables remains unchanged:  $M_{g_js_j}^{(j)} =M_{g_js_j}^{(j-1)} $.

The above predictive distribution  implies that customer $j$ is more  likely to choose   tables with several  occupants, positively reinforcing that table's popularity for future customers. 
 The number of  occupied tables  stochastically increases with the PYP mass  and discount parameters.

 For section $s$, if the PYP discount parameter $d_{s} = 0$, we obtain the well-known P\`{o}lya urn scheme for Dirichlet processes \citep{Ferguson_1973}. 
PYPs act as  effective dimension reduction devices because the random number of occupied tables is much smaller than the number of customers. In general, as the number of  patrons in section $s$ of restaurant $g$ grows as more customers arrive at the franchise, that is, as $p_{gs}^{(j)}\to \infty$,  the number of occupied tables, $M_{gs}^{(j)}$, is asymptotically equivalent to 
\begin{align}
    \begin{cases}
        \alpha_{s}  \log p_{gs}^{(j)}       \qquad &\text{if $d_{s} = 0$} \\
        T_{d_{s} \alpha_{s}} \, p_{gs}^{(j)}\qquad &\text{if $0 < d_{s} < 1$}\\
        \end{cases}\label{q}
\end{align}
for  a positive random variable $T_{d_{s} \alpha_{s}}$   \citep{Lijoi_Prunster_2010}. {Asymptotically, the  order of the number of occupied tables   increases with   discount parameter $d_{s}$.}

\textbf{\textit{Choice of dish $\boldsymbol{\theta}_j$}} \quad  As we have discussed earlier, the 2R2CF process assumes that all  customers seated at a  table of section $s_j$ are served the same dish, chosen from the cuisine $s_j$ menu by the first customer to sit at that table.  
 Let $\boldsymbol{\phi}_{g_j s_j k}$ denote the common dish eaten by the  previous $(j-1)$ customers  at the $k$th table,  $k=1,\ldots,M_{g_js_j}^{(j-1)}$.
 
 Following predictive distribution (\ref{pred.table_j}), 
if customer $j$ has chosen to sit at a previously occupied table, then she or he is served the  dish  selected by the first customer to sit at that table. Otherwise, if customer $j$ has chosen  a new table, then she or he randomly picks a dish from menu distribution $W_{s_j}$. The dish that  customer $j$ eats represents  the probe-specific random effect~$\boldsymbol{\theta}_j$, and
\begin{equation}
    \boldsymbol{\theta}_j  
    \begin{cases}
   = \boldsymbol{\phi}_{g_j s_j v_j}   \qquad &\text{if $v_j = 1,\ldots,M_{g_js_j}^{(j-1)}$,} \\
        \sim W_{s_j}\qquad &\text{if $v_j = (M_{g_js_j}^{(j-1)}+1)$.}\\
    \end{cases}\label{dish}
\end{equation}
In the latter case (line 2), the dish $\boldsymbol{\theta}_j$ randomly selected  by customer $j$ is registered as $\boldsymbol{\phi}_{g_j s_j M_{g_js_j}^{(j)}}$ and is  served to all future customers who sit at  table  $M_{g_js_j}^{(j)}$.  Assumptions~(\ref{pred.table_j}) and (\ref{dish}) imply that although the restaurants  serve  the same  menus, the relative popularity of each dish is  restaurant-specific.

The aforementioned process continues for the remaining 2R2CF customers. Expressions (\ref{PYP_state_prob1}) and (\ref{PYP_grp_prob1}) guarantee that a cuisine is more popular at its namesake restaurant and   the  cuisine selected by a  customer  influences 
the restaurant choice of  the next customer,  
 making the next customer   likely to select the same cuisine. This   accounts for the lengthy runs  of differential or  non-differential  probes observed in methylation data. In addition to achieving dimension reduction, the proposed
Sticky PYP models the  serial dependencies of adjacent probes as a  decreasing function of the inter-probe distances.

\subsubsection{Latent clusters and their differential states}\label{S:latent clusters} Latent clusters, introduced earlier in expression (\ref{allocation c}), comprise  
    probes  with identical random effects and form the basis of the dimension reduction strategy. Returning to the 2R2CF metaphor, we identify a cluster as the set of  customers who eat the same dish. However, in addition to the  customers seated at a table,  multiple tables in both restaurants may  serve  the same dish because of the shared cuisine menu. Therefore, irrespective of the restaurant,
  aggregating  customers eating the same   dishes,  we  obtain   the probe-cluster allocation variables $c_1,\ldots,c_p$, and the  number of latent clusters,  $q$. The sets of customers eating the same cuisine~2 (differential state) dishes correspond to the differential  clusters, $\mathcal{D}$, defined in equation~(\ref{D}).

From expression (\ref{q}), we expect the number of occupied tables to be much smaller than the number of franchise customers,~$p$. Furthermore, since multiple tables may serve the same dish, we expect the number of latent clusters, $q$, to be     smaller than the number of occupied tables.  With high  probability, this implies  that $q$ is much smaller than $p$.

\paragraph{PYP discount parameter $d_2$.}  
Consider  the differential state cuisine menu, $W_2$,  defined in~(\ref{density W2}). It can be  shown that as the number of treatments, $T$, and the number of probes, $p$, increase, 
 the   differential  clusters are not only asymptotically identifiable but  consistently detectable in the posterior; refer to Section 4 of \cite{Guha_Baladandayuthapani_2016} for a detailed general discussion of this remarkable  phenomenon in standard PYP settings. Since the  differential clusters can be   inferred with high accuracy when $T$ and $p$ are large, discount
parameter $d_2$ is  given the mixture prior:
\begin{align*}
d_2\sim \frac{1}{2}\delta_{0} + \frac{1}{2} U(0,1)
\end{align*}
where $d_2=0$  corresponds to a Dirichlet process. 
This provides the posterior flexibility to  choose between a Dirichlet process and a more general~PYP for a suitable clustering pattern of the differential probes.  Allocation patterns  typical of Dirichlet processes, such as exponentially decaying cluster sizes dominated by a few large clusters,  result in high posterior probabilities that $d_2$ equals $0$.
{By contrast,} allocation patterns  characteristic of non-Dirichlet PYPs, {such as slower-than-exponential power law} decays in the cluster sizes and relatively large numbers of smaller-sized clusters,  cause the posterior of discount parameter $d_2$  {to concentrate near 1 and exclude 0.}  A  proof of the intrinsically different cluster patterns  of Dirichlet processes and PYPs is given in  Theorem~2.1 of \cite{Guha_Baladandayuthapani_2016}.

Since distribution $G$ is  discrete, all  atoms of  $T$-variate distribution $W_2$  may not  be  unique. Indeed, this is  common for $T=2$ treatments. However, as $T$ grows, and provided  the number of probes, $p$, grows at a slower-than-exponential rate as $T$, the probability that  two  atoms allocated to the probes are identical rapidly decays to 0. In regression problems  unrelated to differential analysis, Section 2.3 of \cite{Guha_Baladandayuthapani_2016} derived a similar result for a  simpler zero-order stochastic process.  We have verified this phenomenon in  simulation studies on differential analysis datasets. In several hundred  artificial datasets generated from the  Sticky PYP,  for $p=1,500$ probes and  $T$ as small as four, no two allocated atoms of    $W_2$ were  identical.

\paragraph{PYP discount parameter $d_1$.}  Consider the  (non-differential) cuisine menu, $W_1$,  defined in (\ref{W1}).
The   flexibility provided by PYP allocation patterns is not necessary for the  non-differential probes. This is because the allocation patterns of distribution $W_1$  are driven by univariate parameter $\psi$ in (\ref{W1}), and,  in general,  the mixture allocations of univariate objects are unidentifiable \citep[e.g.,][]{Fruhwirth-Schnatter_2006}. Consequently, we set  PYP discount parameter $d_1=0$, reducing the two   PYPs associated with the non-differential state (i.e.,~section 1 in both restaurants) to Dirichlet processes.


\subsubsection{Other model parameters}\label{S:hyperpriors}


  {Depending on the specifics of the application, an appropriate model is assumed for the subject-specific parameters $\xi_1,\ldots,\xi_n$. For example, we may assume that $\xi_i\stackrel{iid}\sim N(0,\tau_{\epsilon}^{2})$. In some  applications, it is more appropriate to assume non-zero means and flexible error distributions: $\xi_i=b_i + \epsilon_i$, where $b_i$ represents lane or batch effects in methylation data, and the i.i.d.\ $\epsilon_i$ have a random distribution with a  Dirichlet process prior whose normal base distribution has mean zero and variance $\tau_{\epsilon}^2$.}
Similarly,  appropriate models  for the probe-specific parameters $\chi_1,\ldots,\chi_p$ may  include i.i.d.~zero-mean normal distributions, and finite mixtures or HMMs with state-specific normal distributions.  Inverse-gamma priors are  assigned to       $\sigma^2$ and $\tau_{\epsilon}^2$.
 Appropriate priors are assumed for   mass parameters $\beta$, $\alpha_1$, and $\alpha_2$  in expressions (\ref{W1}) and (\ref{pred.table_j}). Mean $\mu_G$ and variance $\tau_G^{2}$ of  base distribution $G_0$ in expression (\ref{W1}) are  given a joint normal-inverse gamma~prior. 
Figure \ref{DAG} of Supplementary Materials  displays a directed acyclic graph summarizing the relationships between the different BayesDiff model~parameters.



\section{Posterior Inference} \label{S:post_inf}

Due to the analytical intractability of the BayesDiff model, we rely on MCMC methods  
for posterior inferences  and   detection of  differential probes. 

\subsection{MCMC Strategy} \label{MCMC_summary}


	The  model parameters are  initialized  using na\"{i}ve estimation techniques and   iteratively updated by MCMC techniques until the  chain converges. 
We split the MCMC updates into three  blocks. An outline of the MCMC procedure is as follows. Further details can be found in Section \ref{MCMC_summary_supp} of Supplementary Material.
\begin{enumerate}
    \item\label{MCMC_PYP} \textbf{Restaurant-cuisine-table-dish choice $(g_j,s_j,v_j,\boldsymbol{\theta}_j)$ of customer $j$:} \quad 
    	 For each probe $j=1,\ldots,p,$ we  sample the 4-tuple $(g_j,s_j,v_j,\boldsymbol{\theta}_j)$  given the 4-tuples of the other $(p-1)$ probes. This is achieved by proposing a new value of $(g_j,s_j,v_j,\boldsymbol{\theta}_j)$ from a carefully constructed  approximation to its full conditional, and by accepting or rejecting the proposed parameters in a Metropolis-Hastings step. The procedure is repeated for all $p$ probes.
	 As discussed in Section \ref{S:latent clusters}, the probe-cluster allocations $c_1,\ldots,c_p$  are  immediately available from the restaurant-cuisine-table allocations $(g_j,s_j,v_j)$ of the $p$ probes. Also available are  the $q$ latent clusters along with their allocated probes, and the set of differential clusters $\mathcal{D}$.
    
    \item\label{MCMC_DP} \textbf{Latent vectors} $\boldsymbol{\lambda}_1,\dots, \boldsymbol{\lambda}_q$: \quad There are $T  q$ latent vector elements, not all of which are necessarily distinct because of the Dirichlet process prior on distribution~$G$. Although the latent vector elements are  known from the  Block 1 updates, MCMC mixing is considerably improved by  updating  the latent vector elements conditional on the $p$ probe-cluster allocations. As the  calculation in Supplementary Material shows, this is possible by Gibbs~sampling.

    \item \textbf{Remaining model  parameters}: 
    Generated by standard MCMC techniques. 
\end{enumerate}

For the numerical analyses of this paper, we discarded an initial burn-in  of 10,000 MCMC samples and used the subsequent 50,000 draws for posterior inferences. Convergence was informally assessed by  trace plots of various hyperparameters to validate the MCMC sample sizes. For the proposed moves (in discrete parameter space) described in Step~\ref{MCMC_PYP}, the average Metropolis-Hastings acceptance rate exceeded 90\% in all our analyses.

\subsection{Detection of Differential Probes with FDR Control} \label{FDR_control}
Post-processing the MCMC sample, a Bayesian approach for controlling the false discovery rate (FDR) \citep{newton2004detecting} is applied to accurately detect the   probes $j$ with differential state $s_j=2$.
  Specifically, let $q_0$ be the nominal FDR level and  $\omega_j$ be the posterior probability that probe $j$ is differential, so that $\omega_j=P[s_j=2\mid\boldsymbol{X}]$. An empirical average estimate, $\hat{\omega}_j$, is available from the MCMC sample. To achieve the desired FDR level in calling the differential probes,  we first rank all the probes in decreasing order of $\hat{\omega}_j$. Let $\hat{\omega}_{(1)}>\hat{\omega}_{(2)}>\dots>\hat{\omega}_{(p)}$ denote the ordered posterior probability estimates. For each $b=1,\dots ,p$, we calculate as follows the  posterior expected FDR resulting from calling the first $b$ probes in decreasing order of $\hat{\omega}_j$:
\begin{align}
\widehat{\text{FDR}}_b=\frac{\sum_{j=1}^{p}\left(1-\hat{\omega}_j\right)\mathcal{I}\left(\hat{\omega}_j\geq \hat{\omega}_{(b)}\right)}{\sum_{j=1}^{p}\mathcal{I}\left(\hat{\omega}_j\geq \hat{\omega}_{(b)}\right)}=\frac{\sum_{j=1}^{b}\left(1-\hat{\omega}_{(j)}\right)}{b},
\end{align}
where the simplification occurs because the $\hat{\omega}_j$'s are sorted.
Finally, we pick the largest value of $b$, denoted by $b^{*}$, for which $\widehat{\text{FDR}}_b<q_0$. A nominal FDR~level of $q_0$ is achieved by labeling  as differential the first $b^{*}$ probes arranged in decreasing order of $\hat{\omega}_j$.

\section{Simulation Studies} \label{sim_study}
 
 Using artificial datasets with $T=5$ treatments, we analyzed the accuracy of BayesDiff in detecting  differentially methylated probes. We also  compared the results with  established differential methylation procedures and general statistical techniques for  multiple treatment comparisons. Further, we evaluated  the ability of the BayesDiff procedure in discovering the complex dependence structure of DNA methylation data.

\paragraph{Generation strategy}  \quad Proportions representing DNA methylation data were generated using the logit transformation as in equation (\ref{def_main_model}). 
The inter-probe distances    were the actual distances from the motivating TCGA dataset,   scaled to add to $1$. In order to capture the complexity of methylation data, such as the existence of multiple latent methylation states (e.g,.~CpG
islands and shores),  different read depths across CpGs, and the incomplete
conversion of bisulphite sequencing, the generation strategy was partly based on  techniques  implemented in  WGBSSuite, a flexible stochastic simulation tool for generating single-base resolution methylation data \citep{Rackham_etal_2015}. However, the generation procedure  differed  from WGBSSuite in some respects. Specifically, it allowed more than two treatments ($T>2$). Additionally, as  in actual methylation datasets, the generation procedure     incorporated serial dependence not only in the methylation levels but also in the differential states of the probes.
 
 The probes-specific read depths were generated as $n_j \stackrel{iid}\sim  \text{Poisson}(50)$.
Unlike BayesDiff assumption (\ref{def_main_model}), there were  no subject-specific random effects in the generation mechanism. Instead, the normal mean of the generated data included additive  probe-specific random effects,  $\chi_1^{(0)},\ldots,\chi_p^{(0)}$, that were generated as follows: 

\begin{enumerate}

    \item Generate the true methylation status of the probes, denoted by $h_1^{(0)},\ldots,h_p^{(0)}$, using the 4-state \textit{distance-based} HMM of \cite{Rackham_etal_2015}, with  the states respectively representing the methylated, first transit, demethylated, and second transit states. 
    
    \item Set the baseline methylation levels for the methylated, (first or second) transit, and demethylated states  as $p_{\text{methylated}}=0.8$, $p_{\text{transit}}=0.5$, and  $p_{\text{un-methylated}}=0.2$.

    \item For $j=1,\ldots,p$, define the mean probe-specific random effect as follows:
    \begin{align*}
    \tilde{\chi}_j^{(0)} = 
    \begin{cases}
    \log(\frac{p_{\text{methylated}}}{(1-p_{\text{methylated}}}) &\text{if $h_j=1$ (i.e.,~methylated state),}\\
    \log(\frac{p_{\text{transit}}}{1-p_{\text{transit}}}) &\text{if $h_j=2,4$ (first or second transit state),}\\
    \log(\frac{p_{\text{demethylated}}}{1-p_{\text{demethylated}}}) &\text{if $h_j=3$ (i.e.,~demethylated state).}\\
    \end{cases}
    \end{align*}

    \item Independently generate  true probe-specific random effects: $\chi_j^{(0)} \sim N\bigl(\tilde{\chi}_j^{(0)}, \tau_{\chi}^2\bigr)$ for probe $j=1,\ldots,p$. 

\end{enumerate}

\paragraph{Noise and dependence levels}\quad 
We investigated four  scenarios corresponding to the  combinations of two noise levels and two dependence levels. For each scenario, 20 datasets were independently generated, with each dataset consisting of $p=500$ probes and  $T=5$ treatments with 4 samples each, i.e.,~a total to $n=20$ samples. 
The low noise level corresponded to   true variance parameter  $\sigma^{2}_0=0.36$; equivalently, to  a signal-to-noise of $R_0^2\approx 70\%$. The high noise level corresponded to    $\sigma^{2}_0=1$ or  $R_0^2\approx 40\%$. 
The true between-probe dependencies comprised  two levels: no serial correlation (i.e.,~a zero-order Sticky PYP) with $\eta_0=0$, and positive serial correlation (i.e.,~a first order Sticky PYP) with $\eta_0=0.004$. Although $\eta_0=0.004$ may appear to be small, its value is calibrated to the inter-probe distances;  when the distance between two adjacent probes is equal to the  standardized   average distance of $\bar{e}=1/(p-1)=1/499$, $\eta_0=0.004$ gives an affiliation of $r_0=0.6$ in equation (\ref{def_dep_par}). Since the affiliations are bounded above by 1,  $\eta_0=0.004$  represents fairly high  inter-probe dependence. For convenience, we will refer to the two dependence levels as ``no-correlation'' and ``high correlation.'' The 
other model parameters were common for the four scenarios and are displayed in Table~\ref{parm_sim1}. Setting a true baseline differential proportion of $\rho_2=0.1$ resulted in approximately 10\% true differentially methylated CpGs in each dataset. 

\begin{table}
	\centering
\begin{tabular}{*{9}{c}}
$\alpha_1$  & $\alpha_2$ & $d_2$  & $\beta$ & $\gamma$ & $\rho_2$ & $\mu_G$ & $\tau^{2}_G$ &  $\tau^{2}_{\chi}$  \\  \hline
	20 & 20 & 0.33  & 20 & 0.9 & 0.1 & 0 & 1 &  0.1225
\end{tabular}
\caption{True parameter values used to generate the artificial datasets.}
\label{parm_sim1}
\end{table}

\paragraph{Posterior inferences}  Assuming all model parameters to be unknown, each artificial dataset was analyzed using a BayesDiff model that    differed in key respects from the true generation mechanism. For example, unlike the 4-state HMM of the generation strategy, the probe-specific random effects $\chi_j$ 
 were analyzed  using a BayesDiff model that ignored the  first order dependence, and instead, relied on  a 3-state finite mixture model  representing the methylated, transit, and unmethylated states. Additionally, in contrast to the zeroed-out subject-specific random effects during data generation, the BayesDiff procedure assumed that the random effects were i.i.d.~normal with zero mean.

To assess the accuracy of BayesDiff in detecting the absence or presence of inter-probe serial correlation,  in the no-correlation ($\eta_0=0$) situation, we evaluated 
$\log \left(\frac{P\left[\eta=0\mid \mathbf{X}\right]}{P\left[\eta>0\mid \mathbf{X}\right]}\right)$, the log-Bayes factor comparing  zero order to first order Sticky PYPs. In the high correlation ($\eta_0=004$) situation, we evaluated  $\log \left(\frac{P\left[\eta>0\mid \mathbf{X}\right]}{P\left[\eta=0\mid \mathbf{X}\right]}\right)$, the log-Bayes factor comparing  first order to zero order Sticky PYPs. Thus, in any scenario, a large positive value of this measure  constitutes strong evidence that BayesDiff detects the correct model order.

Although conceptually straightforward, the estimation of Bayes factors requires multiple MCMC runs even for relatively simple parametric models \citep{Chib_1995}. \cite{Basu_Chib_2003} extended the  estimation strategy to infinite dimensional models such as Dirichlet processes. However, the computational costs are prohibitively high for big datasets, and multiple MCMC runs for estimating  Bayes factors  would stretch  available computational resources  beyond their present-day limits. Faced with these challenges, we  relied  on an alternative strategy for estimating the \textit{lower bounds} of log-Bayes factors using a single MCMC run. As it turns out, this is  often sufficient to infer the Sticky PYP model orders.
Let $\Theta^{-}$ denote all BayesDiff model parameters except $\eta$. In the high correlation situation, applying Jensen's inequality, a lower bound for the corresponding log-Bayes factor  is  $E\left[\log \left(\frac{P\left[\eta>0\mid \mathbf{X},\Theta^{-}\right]}{P\left[\eta=0\mid \mathbf{X},\Theta^{-}\right]}\right)\mid\mathbf{X}\right]$. Unlike  log-Bayes factors, this lower bound can be easily estimated by an empirical average estimate based on a single MCMC run. In the no-correlation situation, a lower bound for the  log-Bayes factor, $\log \left(\frac{P\left[\eta=0\mid \mathbf{X}\right]}{P\left[\eta>0\mid \mathbf{X}\right]}\right)$, can be similarly~derived. 

\begin{figure}\centering\includegraphics[width=0.5\textwidth]{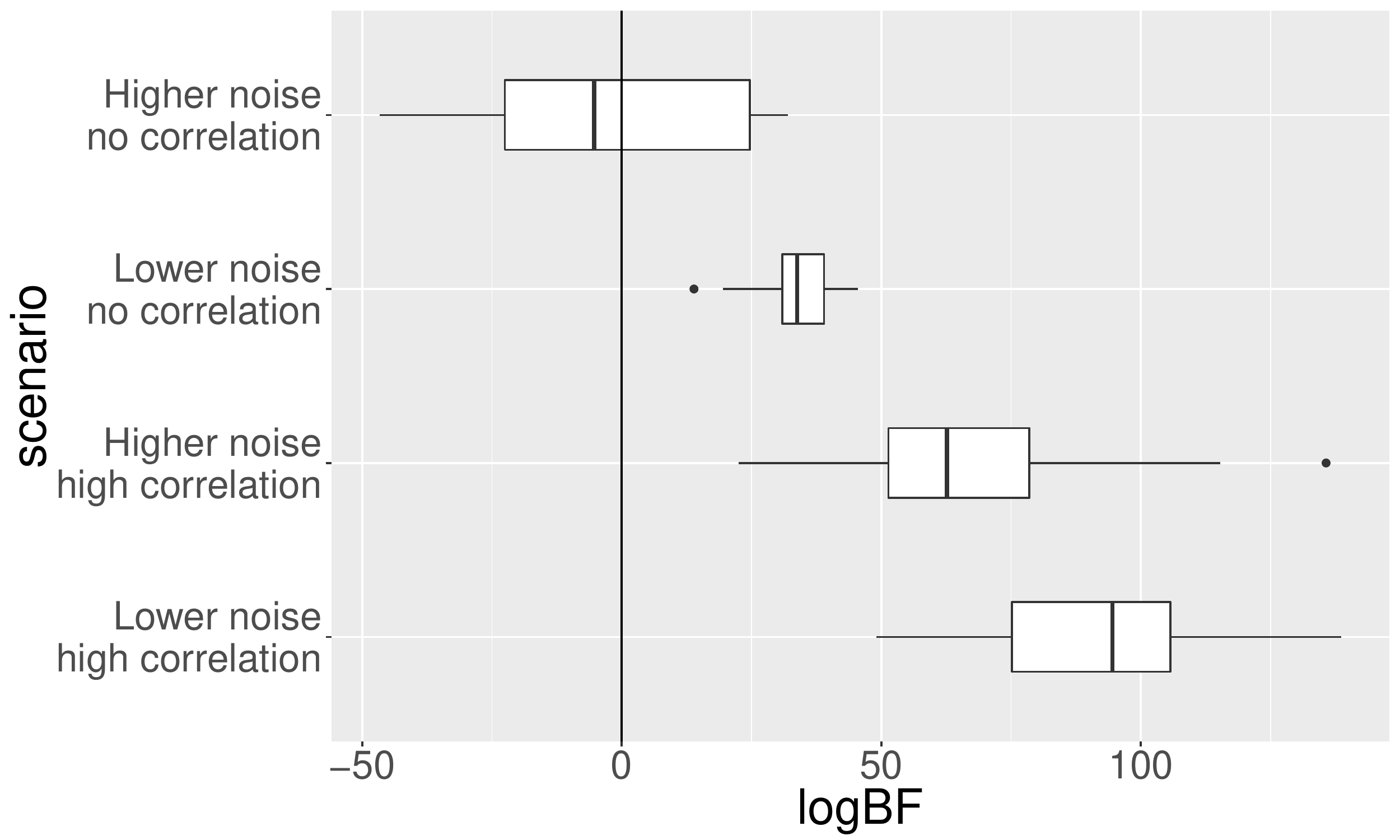}\caption{Simulation study box plots for  estimated lower bounds of  log-Bayes factors in favor of the true model order.}\label{sim1_log-BF}\end{figure}

For each of the four generation scenarios, box plots of these estimated lower bounds for the 20 datasets are depicted in Figure \ref{sim1_log-BF}. Except for the high noise--no-correlation scenario, for which the results were inconclusive, 
the estimated lower bounds of the log-Bayes factors in favor of the true correlation structure were all positive and  large. In the low noise--no-correlation scenario, BayesDiff decisively favored zero-order models, and the smallest lower bound among the 20 datasets was $13.9$,  corresponding to Bayes factors exceeding $e^{13.9}=1,088,161$. The $25^{th}$ percentile of these lower bounds was 30.9,  corresponding to Bayes factors exceeding $e^{30.9}=2.63 \times 10^{13}$.
This is  strong evidence that the BayesDiff approach is reliable in this scenario. For the high-correlation scenarios, the estimated lower bounds were even higher, showing that BayesDiff overwhelmingly favors first order models when the data are actually serially~correlated. 

\paragraph{Comparisons with other methods} 
We evaluated the success of the BayesDiff procedure in detecting  disease genomic signatures and made comparisons with six well-known  procedures. These included some general statistical techniques for multiple treatment comparisons, namely,  one-way analysis of variance (ANOVA) and the Kruskal-Wallis test. We also made comparisons with some methods specially developed  for detecting differential methylation in  more than two treatments: COHCAP \citep{warden2013cohcap}, methylKit \citep{akalin2012methylkit}, BiSeq \citep{hebestreit2013detection}, and RADMeth \citep{dolzhenko2014using}. 
The ANOVA and  Kruskal-Wallis test procedures were applied separately on each probe after applying the inverse-logit transform.  Being specifically designed for differential methylation analysis, the COHCAP method was directly applied to the generated proportions of the synthetic data. The remaining three  methods are designed for bisulfite sequencing, which consists of the total methylation reads for each measured CpG site. For these methods, 
the methylation reads for each probe was obtained by multiplying the proportion methylation values by the total read. The bandwidth smoothing parameter of the method  BiSeq was tuned  to optimize the overall detection.
For all six methods,  probe-specific p-values were evaluated and adjusted for multiplicity using the FDR control procedure of  \cite{benjamini1995controlling}. 

 {As with most   nonparametric Bayes techniques, the associated computational costs of BayesDiff  are considerably higher compared to  frequentist methods.
However, the additional run times are negligible compared to the time frames over
which the experimental data are collected, and as we demonstrate, the trade-off is the substantially
greater accuracy achieved by BayesDiff. On a personal computer with an Intel Core i7-4770 processor with 3.40 GHz frequency  and 8 GB  RAM,  the average computational time for the  Section \ref{MCMC_summary} MCMC algorithm, for the $n=20$ samples, $T=5$ treatments, and $p=500$ probes of the simulation study datasets, was  0.60 seconds per iteration. The computational times were reduced when  different simulation datasets were run in parallel across multiple cores of a research computing cluster.  Analyzing datasets of different sizes,  we find that the computational cost is $O(Tp^2)$ but does not appreciably depend on $n/T$. This is reasonable because the mixture model primarily focuses on  $\boldsymbol{\theta}_1,\ldots,\boldsymbol{\theta}_p \in \mathcal{R}^T$.} 
    {Due to the intensive nature of the  one-parameter-at-a-time Gibbs sampling updates in Block 2, the  Metropolis-Hastings algorithm of \cite{guha2010posterior} can be applied to significantly speed up the updates and make the calculations more scalable. 
As part of ongoing development of  a fast R
package, we find that ten- to hundred-fold speedups are possible with this fast MCMC strategy, which can also be applied   to quickly generate the Block 1 parameters.}

We computed the receiver operating characteristic (ROC) curves for differential probe detection for all seven methods. For a  quantitative assessment, we calculated the area under curve (AUC), declaring the method with the largest AUC  as the most reliable in each scenario.  
The ROC curves, averaged over the 20  datasets under each simulation scenario, are shown with the AUCs in Figure \ref{sim1_ROC_avg} of Supplementary Material. 
In all except the high-noise--no-correlation scenario, BayesDiff uniformly outperformed the other methods. Even in the high-noise--no-correlation scenario,  BayesDiff performed better in the low FPR region.
As expected, all seven methods had lower accuracies for  higher noise levels.
 BayesDiff did significantly better than the competing methods in the high correlation scenarios, suggesting that the incorporation of between-probe dependencies  improves its accuracy in situations typical of DNA methylation data.

Since researchers typically focus on  small false positive rates (FPRs), that is, small significance levels, we also calculated the measures, $\textrm{AUC}_{20}$ and $\textrm{AUC}_{10}$. $\textrm{AUC}_{20}$ ($\textrm{AUC}_{10}$) is defined as the area under the ROC curve  multiplied by $5$ ($10$) when the FPR does not exceed $0.2$ ($0.1$). The multiplicative factor ensures that the areas potentially vary between $0$ and $1$. The three versions of AUC are presented in Table~\ref{table_sim1} in Supplementary Material. 
As also seen in Figure \ref{sim1_ROC_avg},  Table~\ref{table_sim1}   reveals that in three of the four scenarios, BayesDiff had the largest overall AUC. Furthermore, BayesDiff had vastly improved reliability for low FPRs. For example, consider the low noise--high correlation scenario. The overall AUC for BayesDiff was  0.035 greater than that for ANOVA. In contrast, the gains for  BayesDiff, relative to ANOVA, were $+0.107$ for $\textrm{AUC}_{20}$  and $+0.146$ for $\textrm{AUC}_{10}$. The  advantages of BayesDiff were even greater relative to the other competing methods. In the high noise--low-correlation scenario,  BayesDiff had a relatively low AUC, as mentioned. However, even in this scenario, it had the greatest  $\textrm{AUC}_{20}$  and $\textrm{AUC}_{10}$ values among all the methods. Additionally, for a nominal FDR level of $q_0=0.05$, the achieved FDR of BayesDiff  was between 0 and 0.03 in every dataset and  simulation scenario. These results demonstrate the ability of  BayesDiff to accurately detect the differential probes even in challenging situations where the FPR is~small.

\section{Data Analysis} \label{data_analysis}

We returned to the motivating DNA methylation dataset consisting of the upper 
GI cancers: stomach adenocarcinoma (STAD), liver hepatocellular carcinoma (LIHC), esophageal carcinoma (ESCA) and pancreatic adenocarcinoma (PAAD). Applying the BayesDiff procedure, we detected the differentially methylated CpG loci among the cancer types.

\paragraph{Data processing} 
The dataset was obtained from The Cancer Genome Atlas project,  publicly available through The Genomic Data Commons (GDC) portal \citep{grossman2016toward}.
The data are 
available from the Illumina Human Methylation 450 platform for each of 485,577 probes at the CpG sites. At the time of analysis, the dataset consisted of $1,224$ tumor~samples. 
The analysis was performed on a gene-by-gene basis.
 We picked a set of 443 genes  involving mutation in at least 5\% of the samples. 
To ensure that all  CpG sites  potentially related to a gene were included in the analysis, we selected  all    sites  
located
within 50K base pairs outside the gene body, specifically, upstream from the 5' end and downstream from the 3' end. The number of gene-specific CpG sites ranged from 1 to 769, and are displayed in  Figure \ref{appl_plots}(a) of Supplementary Material. As a final preprocessing step, since the methylation patterns of  short genes are  of less interest in cancer~investigations, we eliminated the  25 genes mapped to 20 or fewer CpG sites. 

\paragraph{Inference procedure}\quad
The data were analyzed using the proposed BayesDiff approach. 
Exploratory analyses indicated that a satisfactory fit was obtained by eliminating the probes-specific random effects $\chi_j$  in (\ref{def_main_model}). Since information about  experimental batches is not available for the TCGA dataset, we assumed that the i.i.d.\ $\xi_1,\ldots,\xi_n$ parameters follow a random distribution with a  zero-mean Dirichlet process prior. 
The MCMC procedure of   Section~\ref{MCMC_summary}  was applied to obtain posterior samples for each gene. 
For  detecting  differential  CpG sites, we applied the 
 Section \ref{FDR_control} procedure with a nominal FDR  of $q_0=0.05$.

\paragraph{Results}

\begin{figure}
\centering
\begin{subfigure}{0.4\textwidth}
\centering
\includegraphics[width=1.1\textwidth]{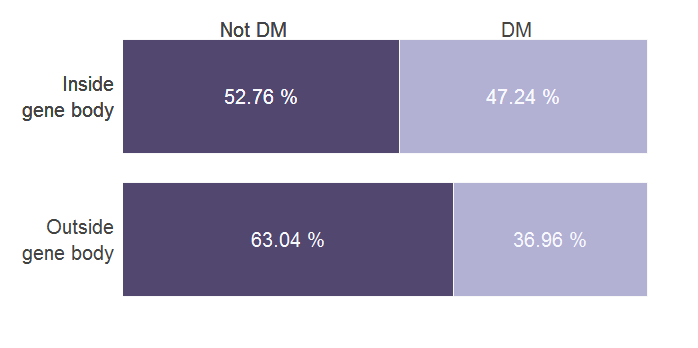}
\caption{Contingency table of detected methylation status and location of CpG site with respect to  gene body}
\end{subfigure}%
\quad
\begin{subfigure}{0.4\textwidth}
\centering
\includegraphics[width=1.1\textwidth]{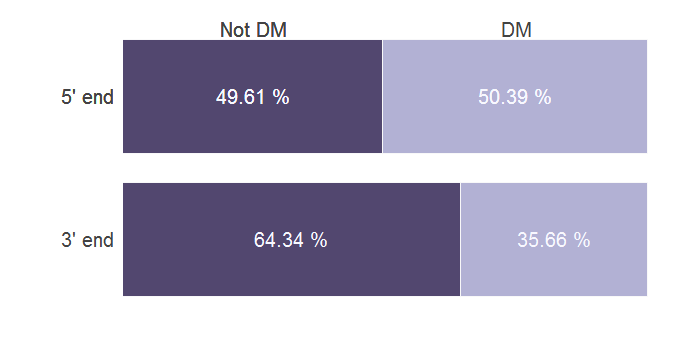}
\caption{Contingency table of detected methylation status and proximity  of CpG site to  chromosomal end}
\end{subfigure}
	\caption{Associations of  detected methylation status and position of  CpG sites. }
	\label{appl_probeSummaries}
\end{figure}

Among the differentially methylated CpG sites detected by our approach, 
approximately 40.6\% of the sites were located outside the gene bodies. 
Figure \ref{appl_probeSummaries} displays the  associations between  detected methylation status and positions of the CpG sites. 
For our analysis, we have defined ``near the 5' (3') end'' as the CpG sites located within one-fourth length of the gene body, either inside or outside the gene boundary, and closer to the transcription start (termination) site.
Our results indicate  that the proportion of differential methylation is higher for CpG sites inside the gene body, and that most differentially methylated loci are situated within the gene body, as is well known from numerous previous studies. However, our analysis also revealed significant amounts of differential methylation outside the gene body. Despite the common belief that DNA methylation analysis should focus on the 5' end region, we  found that CpG sites near the 3' ends also displayed  considerable degrees of differential methylation. These findings support the recommendations of \cite{irizarry2009genome}  that   studies of  DNA methylation alteration 
should be conducted on a  higher resolution, epigenome-wide basis.

\begin{figure}
\centering
\includegraphics[width=0.9\textwidth]{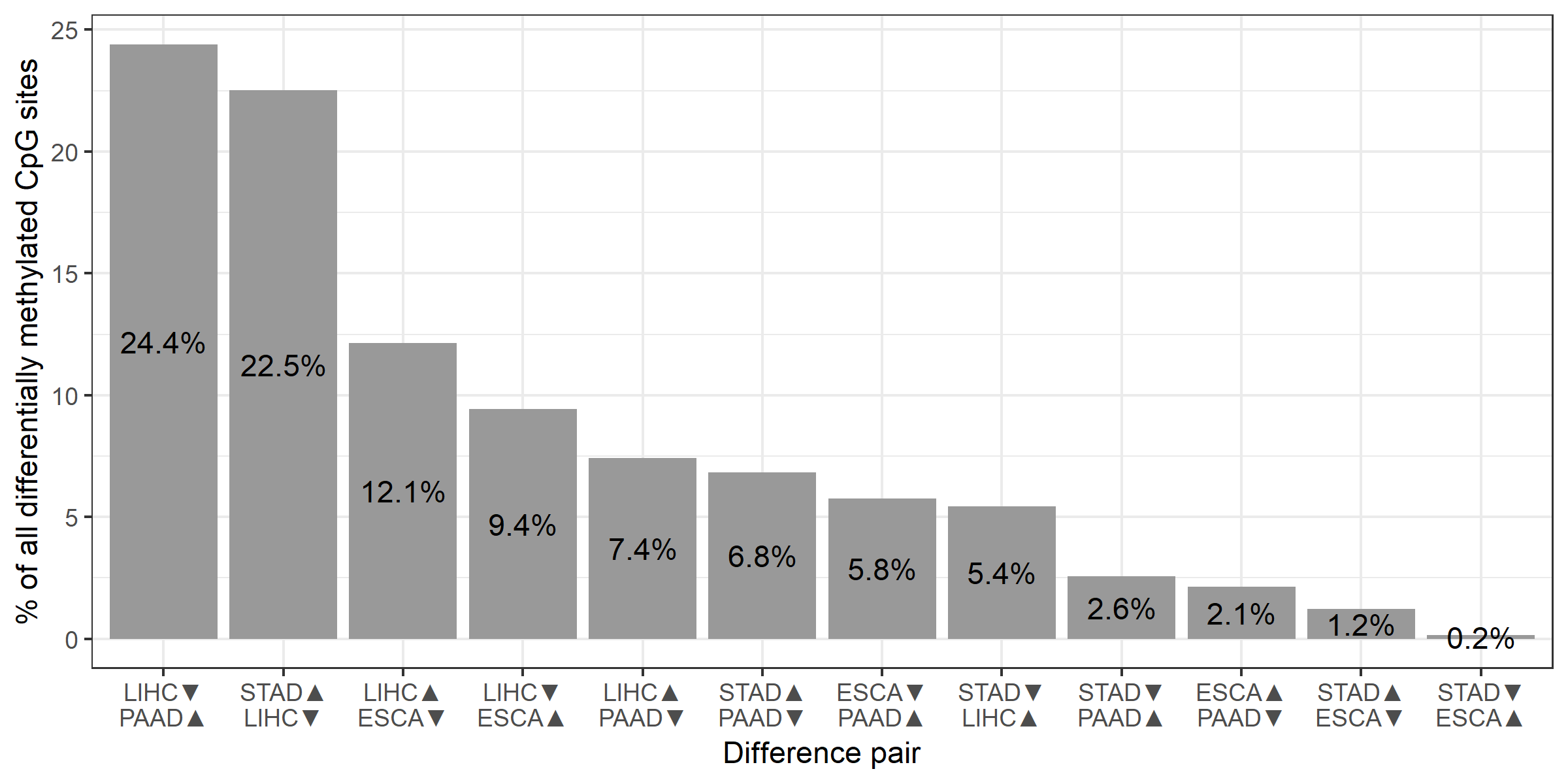}
	\caption{Site-wise summary of the largest pairwise differences of differentially methylated loci among the four upper GI cancer types.}
	\label{appl_maxdifSummaries}
\end{figure}

Among the  differentially methylated sites detected by BayesDiff, we estimated the pairwise  differences between  random effects associated with the four cancer types. Site-wise
 summaries of the    largest pairwise differences of  the cancer-specific effects are displayed in Figure \ref{appl_maxdifSummaries}. None of the four cancer types displayed consistent hypermethylation or hypomethylation across all  genes or   entire chromosomes. However, we found that LIHC is frequently   differentially methylated relative to one of the other cancer types, implying that it is the most volatile disease  with respect to DNA~methylation.

For each gene, Figure \ref{appl_plots}(b) of Supplementary Material displays 95\% credible intervals for  lower bounds of log-Bayes factors of a first versus zero-order model, i.e., $\eta=0$ versus $\eta > 0$ in expression (\ref{def_dep_par}). 
Models with first order dependence are overwhelmingly favored for a majority of the genes, suggesting that statistical techniques that  do not account for dependence between neighboring  CpG sites would be less effective for these data. 
Figure~\ref{DA_detail_2genes} of Supplementary Material  displays the detailed differential methylation pattern  for the top two mutated genes, TP53 and TTN. An obvious feature of both the genes is that the detected differential methylation of the CpG sites is highly serially correlated. For gene TP53, there are almost no differentially methylated loci  within the gene body. The 3' end region outside the gene body has a cluster of differentially methylated loci, for which cancer type STAD is  mostly hypermethylated. The results for gene TTN tell a quite different story: most of the differentially methylated loci are inside the gene body and near the 5' end. Cancer type LIHC is hypomethylated compared to PAAD around the 5' end region, but it is hypermethylated compared to STAD near the 3' end. 
Genes with at least 90\% differentially methylated sites detected by BayesDiff are listed in Table \ref{table_list_gene_prop} of Supplementary Material, along with 
the largest pairwise difference between the four cancer types among the differentially methylated loci.
The  number of CpG sites within each segment is listed in Table~\ref{table_list_num_probe} of Supplementary Material.

Existing medical literature both supports and complements our findings.  For example, hypermethylation of the EDNRB and SLIT2 genes have been found in STAD \citep{tao2012quantitative}. 
 Gene FBN2 was  hypermethylated in ESCA \citep{tsunoda2009methylation}. 
 While several studies have found that the gene and protein expressions of ABC transporter genes, such as ABCC9, are useful for understanding the prognosis of esophageal cancer \citep{vrana2018abc}, we conclude that  hypermethylation of ABCC9 is a major difference between   cancer types ESCA and LIHC. Gene FLRT2 is a potential tumor suppressor   that is hypermethylated and downregulated in breast cancer \citep{bae2017epigenetically}. Our results indicate that this gene is also hypermethylated in  cancer type STAD versus LIHC. Mutations in SPTA1 gene has been linked with PAAD \citep{murphy2013genetic}.  Our results  indicate that  hypermethylation of this gene distinguishes PAAD from LIHC.

{Finally,  we compared our findings with those of ANOVA  for multiple treatment comparisons. Table~\ref{table_list_overlap_gene_prop} lists the common set of genes with at least 90\% differentially methylated
sites identified by both BayesDiff \textit{and} ANOVA. Table~\ref{table_list_ANOVA_only_gene_prop} displays the genes  identified by \textit{only} ANOVA, whereas   Table~\ref{table_list_BayesDiff_only_gene_prop} displays the large number of genes  detected exclusively by BayesDiff. 
Cross-referencing  with the medical literature, we find that  genes FLRT2 and  FBN2 were detected by both methods. However, genes  EDNRB, SLIT2, ABCC9, and SPTA1 were only identified by BayesDiff,     revealing the benefits of  the proposed method.}

\paragraph{Accounting for data characteristics}  
A  statistical model for biomarker data should account  for the observed probe-specific means and variances. 
This is especially important in multiple-testing based approaches where  the first two sample moments  must be plausibly explained by the fitted model   to avoid making misleading biological interpretations \citep{Subramaniam_Hsiao_2012}. 
From this perspective, certain aspects of the 
 BayesDiff model,  such as 
  variance~$\sigma^2$ being a priori unrelated to the mean in expression~(\ref{def_main_model}), may  appear to be  restrictive. However, even though the BayesDiff model was not specifically constructed to match data summaries such as sample moments,  
in practice, the nonparametric nature of the Sticky PYP  
allows the  posterior to flexibly adapt to the features of the data, including   sample moments, thereby  accounting for   mean-variance relationships in a robust manner. 
For example, consider again the top  mutated genes, TP53 and TTN, discussed in Figure \ref{DA_detail_2genes} of Supplementary Material. The ability of the BayesDiff model to match the sample moments of the gene-specific probes is demonstrated as follows. 
Given the inter-probe distances, the joint posterior of the BayesDiff parameters induces predictive distributions on the $n$ measurements for each probe. Functionals of these predictive distributions, such as the probe-specific sample moments, are easily estimated by post-processing the MCMC sample. For these two genes, Figure \ref{mean-mean_var-var} of Supplementary Material shows that the  sample moments predicted by BayesDiff closely match the actual first and second sample moments, with  correlations exceeding 99\% in each plot. Similar results were observed in other datasets.

\section{Discussion} \label{disc}

DNA methylation data  exhibit complex  structures due to unknown biological mechanisms and distance-dependent serial correlations among neighboring CpG sites or probes. The identification of differential  signatures among multiple treatments or sets  of tumor samples 
is crucial for developing targeted treatments for disease.
This paper formulates a flexible  approach applicable to multiple treatments called BayesDiff. The technique relies on a novel Bayesian mixture model called the Sticky PYP or  the two-restaurant two-cuisine franchise.  In addition to  allowing  simultaneous inferences on the probes, the model accommodates distance-based serial dependence and 
accounts for the complex interaction patterns commonly observed in cancer data. A effective MCMC strategy  for  detecting the differential probes is developed.
The success of  the BayesDiff procedure in differential DNA methylation, relative to well-established methodologies,  is  exhibited via simulation studies. The new technique is applied to the motivating TCGA dataset to detect the  differential genomic signatures of four upper GI~cancers.
The results both support and complement various known facts about the epigenomic differences between these cancer types, while   revealing  a set of genes exhibiting high proportions of differentially methylated CpG sites. 

In addition to providing a good fit for the data, a  statistical model must be able  to account  for  features of the dataset such as sample moments. The success of the BayesDiff model in this regard is demonstrated in  Section \ref{data_analysis} for the upper GI dataset.
It must be emphasized, however, that there may be   datasets where BayesDiff is less successful in accounting for the data characteristics, possibly  due to  slow asymptotic  convergence of the posterior to the true generative process. 
In such situations, more flexible global transformations \citep{Li_etal_2016} or variance-stabilizing transformations \citep{Durbin_etal_2002} may be used.  Alternatively, 
 local Laplace approximations of exponential family likelihoods through link functions 
  \citep{Zeger_Karim_1991, Chib_Winkelmann_2001}  
 may  extend the BayesDiff procedure 
  and better explain  unique data characteristics.  

{Like most   Bayesian models comprising several latent parameters, the proposed 2R2CF can  be marginalized over different parameter sets to obtain equivalent  versions of the same model. 
 For example, we could marginalize over the \textit{restaurants} to obtain an equivalent ``sticky cuisine'' version in which there is just  one restaurant with two cuisine-sections and  a customer more likely to favor the \textit{cuisine} selected by the previous customer. 
Alternatively, we could integrate out the \textit{sections} to obtain an equivalent ``sticky restaurant franchise'' in which each restaurant comprises  a single  section with  restaurant-specific probabilities that guarantee that Cuisine 1 or 2 dishes are more popular at their namesake restaurant; a customer  is then more likely to favor the \textit{restaurant} selected by the previous customer. 
In all equivalent versions, however,  a probe's differential state 
is determined by the characteristics of the  customer's dish in the~metaphor.}

 {The 2R2CF perspective offers the twin advantages of parameter interpretability and  generalizability. Section \ref{general2R2CF} of Supplementary Material presents the   generalized form of the Sticky PYP, revealing the full potential of the proposed method in analyzing not only DNA methylation datasets, but also  other types of omics datasets, such as gene expression, RNASeq, and copy-number alteration data.  Beyond biomedical applications, 
 the generalized formulation offers a diverse palette of parametric and nonparametric
models for capturing the distinctive features of datasets. These Bayesian  mixture models are special cases of Sticky PYPs for particular choices of a countable group parameter (e.g., two ``restaurants'' in the 2R2CF metaphor for differential methylation problems) and countable state parameter (e.g., two ``cuisines'' in  2R2CF) with the state of a customer influencing the  group of the next customer.   In addition to extending PYPs to discrete time series type data, the range of  models
includes Dirichlet processes, PYPs, infinite HMMs, hierarchical Dirichlet processes \citep{Teh_etal_2006}, hierarchical Pitman-Yor processes \citep{Teh_etal_2006,camerlenghi2019distribution}, finite HMMs, nested Chinese restaurant processes \citep{Blei_Jordan_2005}, nested Dirichlet processes \citep{Rodriguez_etal_2008}, and
analysis of densities models \citep{Tomlinson_Escobar_2003}. }

Ongoing work  involves extending the  correlation structure  to model more sophisticated forms of inter-probe dependence in DNA methylation data. 
{Commented R
code implementing the BayesDiff method is  available on GitHub at https://github.com/cgz59/ BayesDiff.} Using
high-performance Rcpp subroutines,
we are  developing a fast R package  for  detecting differential genomic signatures in a wide variety of omics datasets. Initial results indicate that  dramatic  speedups  of several orders of magnitude would 
allow the fast analyses of high-dimensional datasets  on  personal
 computers.


\newpage


\begin{center}
{\large\bf SUPPLEMENTARY MATERIAL}
\end{center}

\section{MCMC Strategy} \label{MCMC_summary_supp}


	As mentioned in Section \ref{MCMC_summary} of the paper, 
we split the MCMC updates into three  blocks. The MCMC procedure for updating Blocks 1 and 2 is described below. Unless otherwise stated, all references to equations, tables, and figures are for the main paper.
\begin{enumerate}
    \item \textbf{Restaurant-cuisine-table-dish choice $(g_j,s_j,v_j,\boldsymbol{\theta}_j)$ of customer $j$:} \quad 
    	 For each probe $j=1,\ldots,p,$ we  sample the 4-tuple $(g_j,s_j,v_j,\boldsymbol{\theta}_j)$  given the 4-tuples of the other $(p-1)$ probes. This is achieved by proposing a new value of $(g_j,s_j,v_j,\boldsymbol{\theta}_j)$ from a carefully constructed  approximation to its full conditional, and by accepting or rejecting the proposed 4-tuple of probe-specific parameters in a Metropolis-Hastings step. The procedure is repeated for all $p$ probes to complete  the block of MCMC updates.

	We give here an intuitive description of the  details. Further calculation details are available in Section \ref{MCMC_sec} of Supplementary Material. For the $j$th probe, $1<j<p$, the choice of restaurant $g_j$ depends on the triplets of the two immediately adjacent probes. Specifically, as discussed in Section~\ref{Sticky_PYPs_DA} and  graphically shown in the upper and middle panel of Figure \ref{2R2CF plots}, the restaurant selected by a customer depends on the cuisine  of the previous customer. 
	Then, within restaurant $g_j$, the customer chooses cuisine $s_j$ with a restaurant-specific probability. Finally, the customer may either sit an existing table, in which case they must eat the same dish as everyone already sitting at that table, or the customer may open a new table and order a dish from cuisine menu $s_j$. The generation strategy for  table and dish depends on the cuisine, as discussed below. 

	\vspace{-5 mm}
	
	\paragraph{Cuisine $s_j=1$} Evaluating the posterior probability of  table  $v_j$ involves integrating the Gaussian likelihood  of column vector $\mathbf{z}_j=(z_{1j},\ldots,z_{nj})'$ with respect to the cuisine~1 menu, i.e.,~$\boldsymbol{\theta}_j
        \sim W_{1}$. Recall that the cuisine 1  menu has the special structure (\ref{W1}),  allowing its reduction to a univariate quantity distributed as $G$. This random distribution itself  follows  a Dirichlet process conditional prior  with a base distribution that is a mixture of known atoms (given the 4-tuples of the other probes) and Gaussian distribution $G_0$. This conjugate hierarchical structure allows the calculation of  table and dish choice for cuisine 1 in computationally closed~form.
	
		\vspace{-5 mm}

\paragraph{Cuisine $s_j=2$} This situation is more complicated 
because 
 the Gaussian likelihood must be integrated with respect to  prior distribution $W_{2}$ of the cuisine~2 dishes. Unlike cuisine 1, this is  not possible in computationally closed form because menu structure (\ref{density W2}) cannot be reduced to a univariate quantity. 
        In other words, for cuisine 2 (i.e.,~the differential state), the Dirichlet process conditional prior for distribution $G$  is non-conjugate.

We utilize an auxiliary variable approach. Given the current set and frequency of atoms of distribution $G$ in the other $(p-1)$ probes, we first compute a finite mixture approximation, $G^{*}$, to the Dirichlet process conditional prior for  $G$, as in \cite{Ishwaran_Zarepour_2002}. Then, using prior (\ref{density W2}), it is possible to  approximate the posterior probability of customer $j$  joining an existing table or sitting at a new table, irrespective of the dishes. This allows us to generate the table selection, $v_j$. Finally, we  propose the dish, $\boldsymbol{\theta}_j=\boldsymbol{\phi}_{g_j s_j v_j}$. If  table $v_j$ is one of the already existing tables, then the customer simply chooses the common, table-specific dish. Otherwise, we sequentially generate from the posterior the $T$ elements representing the new dish. This is done  using the corresponding treatment-specific elements of vector $\mathbf{z}_j=(z_{1j},\ldots,z_{nj})'$ and finite mixture $G^{*}$, with the restriction that not all $T$ atoms of vector $\boldsymbol{\theta}_j$ are  equal. The Metropolis-Hasting step compensates for any approximations and produces post--burn-in samples from the BayesDiff model posterior.

		\vspace{-5 mm}
		
   \paragraph{Latent clusters} As discussed in Section \ref{S:latent clusters}, the probe-cluster allocations $c_1,\ldots,c_p$  are  immediately available from the restaurant-cuisine-table allocations $(g_j,s_j,v_j)$ of the $p$ probes. The $q$ latent clusters, along with their allocated probes, and the set of differential clusters $\mathcal{D}$, are also immediately available.
    
    \item \textbf{Latent vectors} $\boldsymbol{\lambda}_1,\dots, \boldsymbol{\lambda}_q$: \quad  Since each  latent vector  is of length $T$,  there are $T  q$ latent vector elements, not all of which are necessarily distinct because of the Dirichlet process prior on distribution $G$. Although the latent vector elements are  available from  Block 1, mixing of the MCMC chain is considerably improved by an additional block of updates of the latent vector elements conditional on the $p$ probe-cluster allocations. As the following calculation shows, this can be done by Gibbs~sampling.  
    
    Since a non-differential  cluster's  latent vector consists of a single  value repeated $T$ times, the number of \textit{distinct} latent vector elements  is no more than $\left(q_1+T  q_2\right)$.   Given the  current probe-cluster allocations, suppose there are $m_k$ probes associated with latent cluster $k$, where $k=1,\dots, q$.  Let $q_2=|\mathcal{D}|$ be the  number of differential clusters, so that $q_1=q-q_2$ is the  number of non-differential clusters.  
    
    	\vspace{-5 mm}
    	
    \paragraph{Non-differential clusters}
    For these clusters, the latent vector $\boldsymbol{\lambda}_k$ takes the form, $\boldsymbol{\lambda}_k = \psi_k \boldsymbol{1}_T$ for some $\psi_k \in \mathcal{R}$. Given the  other $(q-1)$  latent vectors, assumptions (\ref{W1}) and (\ref{density W2})  imply that   parameter $\psi_k$  has a Dirichlet process conditional prior. The  base distribution of the Dirichlet process is a mixture of a continuous distribution, $G_0=N(\mu_G, \tau_G^2)$, and the univariate atoms located at  the ``known''   elements of the other $(q-1)$  latent vectors. The sufficient statistic for    $\psi_k$  is	\begin{align*}
		\bar{y}_{ k} &= \frac{1}{n m_k}\sum_{j=1}^{p}\sum_{i=1}^{n}\left({z_{ij}}-{\xi_i}-\chi_j\right)\mathcal{I}\left(c_j=k\right) \\&\sim N\left(\psi_k,\frac{\sigma^2}{n m_k}\right) \quad\text{for cluster } k \not\in \mathcal{D}.
	\end{align*}
Since the  Dirichlet process conditional prior 
	is conjugate to this normal likelihood, parameter $\psi_k$  could be updated by Gibbs sampling \citep{escobar1994estimating,maceachern1994estimating,escobar1995bayesian}.
	
		\vspace{-5 mm}
		
	 \paragraph{Differential clusters} For these clusters,  some of  the $T$ elements of
  latent vector $\boldsymbol{\lambda}_k = (\lambda_{1k},\ldots,\lambda_{Tk})'$ could be    tied, but at least two elements must be unequal. Denote these restrictions on $\boldsymbol{\lambda}_k$ by $\mathcal{A}$. For a  treatment~$t$,  let the remaining $(T-1)$ elements of vector $\boldsymbol{\lambda}_k$ be denoted by $\boldsymbol{\lambda}_{-tk}$.
  
 Imagine updating element $\lambda_{tk}$ assuming that $\boldsymbol{\lambda}_{-tk}$ and the remaining $(q-1)$ latent vectors are  known.  Vector $\boldsymbol{\lambda}_{-tk}$ and  restriction $\mathcal{A}$ imply a possibly restricted support, $\mathcal{A}_t$, for parameter $\lambda_{tk}$.  Observe that the support is unrestricted, i.e.,~$\mathcal{A}_t=\mathcal{R}$, if at least two  elements of  $\boldsymbol{\lambda}_{-tk}$ differ.
  Similarly  to non-differential clusters, it can be shown that    element $\lambda_{tk}$  follows a Dirichlet process conditional prior restricted to  the support $\mathcal{A}_t$, and that the  base distribution is a  mixture of a  normal distribution  and  univariate atoms. The  sufficient statistic for latent vector element $\lambda_{tk}$ is
	\begin{align*}
		\bar{y}_{t k}&=\frac{1}{n_t m_k} \sum_{j=1}^{p}\sum_{i:t_i=t}\left({z_{ij}}-{\xi_i}-\chi_j\right)\mathcal{I}\left(c_j=k\right) \\
		&\sim N\left(\lambda_{t k},\frac{\sigma^2}{n_t m_k}\right) \quad\text{for treatment   $t=1,\ldots,T$, and cluster $k \in \mathcal{D}$},
	\end{align*}
	where $n_{t}$ is the number of individuals associated with treatment $t$. The conjugate structure  implies that parameter $\lambda_{tk}$ could be generated from its full conditional  by a rejection sampler on  set $\mathcal{A}_t$ coupled with standard Gibbs proposals for conjugate 
	 Dirichlet processes. 
	  In practice, the acceptance rates of the rejection sampler are very  high  for $T$ as small as 3 or 4, and are nearly 100\% for larger  $T$. This is because the posterior probability of  support set $\mathcal{A}_t$ tends to  1 as $T$ grows. An intuitive explanation for this asymptotic property is provided in Section \ref{S:hyperpriors} following the prior specification of discount parameter~$d_2$.

\end{enumerate}

	\section{Additional details about Block 1 MCMC updates} \label{MCMC_sec}

	In the first block of  MCMC updates of  Section \ref{MCMC_summary}, we sample the the set \[\mathcal{C} = \left\{(g_j,s_j,v_j,\boldsymbol{\theta}_j) : j = 1,\ldots, p\right\}.
	\]
	We iteratively sample the 4-tuple, $(g_j,s_j,v_j,\boldsymbol{\theta}_j)$, for each probe $j$ via Metropolis-Hastings updates. 
	Specifically, for the $j$th probe, we  propose group (i.e.,~restaurant) $g_j$, state (i.e.,~cuisine) $s_j$, PYP cluster (i.e.,~table) label $v_j$, and random effect (i.e.,~dish) $\boldsymbol{\theta}_j$, with the proposal distribution approximately equal to the joint posterior  of $(g_j,s_j,v_j,\boldsymbol{\theta}_j)$ conditional on the data and   all other parameters being equal to the current values. As explained in the paper, since the base measure of the Sticky PYP  is discrete and not conjugate to the likelihood, this makes the case of parameter $v_j$ being assigned a new cluster label  complicated. To deal with this situation, an auxiliary variable approach is used. Conditional on the current parameter values of the Dirichlet process prior for the latent vector elements, we generate a finite-dimensional approximation, $G^{*}$, to the Dirichlet process \citep{Ishwaran_Zarepour_2002}. Since the distribution $G^{*}$ is discrete, it is  characterized  by a finite vector of probability masses, $\boldsymbol{\pi}$, and the corresponding values of probability mass points, $\mathbf{u}$. Then, conditional on the auxiliary variables from $G^{*}$, we obtain the quantities needed for the conditional posterior of $v_j$ being assigned a new cluster label in the Metropolis-Hastings algorithm. 
	
	More formally, consider probe $j \in \{1,\dots ,p\}$. The proposal probabilities for $\left(g_j,s_j,v_j\right)$ are as follows:
	\begin{align} 
		&\bullet \quad \text{For an existing PYP cluster indexed by } v\in \{1,\ldots,q_{gs}^{(-j)}\},\nonumber \\
		&Q\left(g_j=g,s_j=s,v_j=v\right)=P\left(g_j=g,s_j=s,v_j=v\mid \mathbf{z}_j,\mathbf{c}^{(-j)},\boldsymbol{\phi}_{gsv},s_{j-1},\eta,\rho,\gamma,\boldsymbol{\xi},\boldsymbol{\chi},\sigma^{2}\right) \nonumber \\
		&\propto \left[\mathbf{z}_j\mid v,s,g,\boldsymbol{\phi}_{gsv},\boldsymbol{\xi},\boldsymbol{\chi},\sigma^{2}\right]P\left(v_j=v\mid \mathbf{c}^{(-j)},s,g\right)P\left(s_j=s\mid g,\gamma,\rho\right)P\left(g_j=g\mid s_{j-1},\eta,\gamma,\rho\right) \nonumber \\
		&\propto \prod_{i=1}^{n}\varphi\left({z_{ij}}\mid \phi_{t_igsv}+\xi_i+\chi_j,\sigma^{2}\right)\left(m^{(-j)}_{gsv}-d_{s}\right) \mathcal{Q}_g(s) \mathcal{F}_{j}(g).
		\label{proposal_existing}
	\end{align}
	\begin{align}
		&\bullet \quad\text{For a new PYP cluster indexed by } v^{*}= q_{gs}^{(-j)}+1, \nonumber \\
		&Q\left(g_j=g,s_j=s,v_j=v^{*}\right)=P\left(g_j=g,s_j=s,v_j=v^{*}\mid \mathbf{z}_j,\mathbf{c}^{(-j)},\boldsymbol{\pi},\mathbf{u},s_{j-1},\eta,\rho,\gamma,\boldsymbol{\xi},\boldsymbol{\chi},\sigma^{2}\right) \nonumber \\ 
		&\propto
		\left[\mathbf{z}_j\mid v^{*},s,g,\boldsymbol{\pi},\mathbf{u},\boldsymbol{\xi},\boldsymbol{\chi},\sigma^{2}\right]P\left(v_j=v^{*}\mid \mathbf{c}^{(-j)},s,g\right)P\left(s_j=s\mid g,\gamma,\rho\right)P\left(g_j=g\mid s_{j-1},\eta,\gamma,\rho\right) \nonumber \\
		&\propto 
		\begin{cases}
			\!\begin{aligned}
				& \sum_{l=1}^{L}\pi_l\left\{\prod_{i=1}^{n}\varphi\left(z_{ij}\mid u_l+\xi_i+\chi_j,\sigma^{2}\right)\right\}\left(\alpha_1+d_{1}q^{(-j)}_{g1}\right) \mathcal{Q}_g(1) \mathcal{F}_{j}(g),  \quad \text{if $s=1$,} \\
			\end{aligned}            \\
			\!\begin{aligned}
				\left\{\prod_{t=1}^{T}\sum_{l=1}^{L} \pi_l \prod_{i:t_i=t}\varphi\left({z_{ij}}\mid u_l+{\xi_i}+\chi_j,\sigma^{2}\right) 
				-\sum_{l=1}^{L} \pi_{l}^{T} \prod_{i=1}^{n}  \varphi\left({z_{ij}}\mid u_{l}+{\xi_i},\sigma^{2}\right)\right\} \\
				\times \left(\alpha_2+d_{2}q^{(-j)}_{g2}\right) \mathcal{Q}_g(2) \mathcal{F}_{j}(g),  \quad\text{if $s=2$.} \\
			\end{aligned}            \\
		\end{cases} 
		\label{proposal_new}
	\end{align}
	In expressions (\ref{proposal_existing}) and (\ref{proposal_new}), when $j<p$,  superscript $(-j)$ for a variable indicates that the  calculation excludes the $j$th and $(j+1)$th cluster allocation variables. Specifically, $q^{(-j)}_{gs}$ denotes the number of PYP clusters for group $g$ and state $s$; $\mathbf{c}^{(-j)}$ denotes the vector of  cluster allocations; $m^{(-j)}_{gsv}$ denotes the cluster membership count for cluster $v$ in the PYP with group $g$ and state $s$; $\mathcal{Q}_g(s)$ and $\mathcal{F}_{j}(g)$ are defined in the paper; $\varphi\left(\cdot\mid \mu,\sigma^{2}\right)$ is the density of the normal distribution with mean $\mu$ and variance $\sigma^{2}$; $L$ is the number of distinct mass points in $G^{*}$; $u_l$ denotes the value of distinct mass points in $G^{*}$ and $\pi_l$ is the corresponding probability mass. Before calculating the proposal, latent vectors of emptied clusters are removed and cluster labels  rearranged so that the largest label is~$q^{(-j)}_{gs}$. 

	The case
	$v^{*}= (q_{gs}^{(-j)}+1)$ corresponds to the case of $v_j$ opening a new cluster. If $v_j=(q_{gs}^{(-j)}+1)$ has been proposed, then a new random effect of length $T$ is sampled from the posterior distribution based on the $n$-dimensional  observation vector $\mathbf{z}_j$ and the prior $G^{*}$ characterized by the auxiliary variables, subject to the constraints imposed on the random effects of the differential state $s_j$. 
	
	The Metropolis-Hastings acceptance ratio is  computed to decide whether the proposed $g_j$, $s_j$ and $v_j$ values are accepted. For $j=p$, our proposal density is exactly the desired conditional posterior, so we  always accept the move. For $j<p$, since our proposal density is part of the desired conditional posterior for $\left(g_j, s_j, v_j\right)$, this part cancels out in the acceptance ratio. Therefore, the acceptance ratio  only relates to the transition from $\left(g_j, s_j, v_j\right)$ to $\left(g_{j+1}, s_{j+1}, v_{j+1}\right)$, which is
	\begin{align}
		r_j=\frac{p\left(g_{j+1},s_{j+1},\tilde{v}_{j+1},\mid g_j^{*},s_j^{*},v_j^{*},\boldsymbol{\phi}^{*}_{g_{j+1}s_{j+1}\tilde{v}_{j+1}},\mathbf{z}_{j+1},\mathbf{c}^{-},\eta,\boldsymbol{\xi},\boldsymbol{\chi},\sigma\right)}{p\left(g_{j+1},s_{j+1},v_{j+1} \mid g_j^{0},s_j^{0},v_j^{0},\boldsymbol{\phi}_{g_{j+1}s_{j+1}v_{j+1}},\mathbf{z}_{j+1},\mathbf{c}^{-},\eta,\boldsymbol{\xi},\boldsymbol{\chi},\sigma\right)}
		\label{MH_ratio}
	\end{align}
	where $g_j^{*}$, $s_j^{*}$ and $v_j^{*}$ represent the proposed values;  $g_j^{0}$, $s_j^{0}$ and $v_j^{0}$ are the old values; $\tilde{v}_{j+1}$ denotes  $v_{j+1}$ under the proposed variable values (due to possible PYP cluster label change or elimination). Specifically, $\tilde{v}_{j+1}=(q^{*}_{g_{j+1}s_{j+1}}+1)$ if it belongs to no existing cluster, where $q^{*}_{g_{j+1}s_{j+1}}$ is the number of clusters under the proposed variable values and $\boldsymbol{\phi}^{*}_{g_{j+1}s_{j+1}\tilde{v}_{j+1}}$ is the random effect under the proposed variable values. Both the nominator and the denominator in (\ref{MH_ratio}) can be calculated using (\ref{proposal_existing}) and (\ref{proposal_new}), and by transposing $j$ to $(j+1)$.

\section{Generalized form of the  Sticky PYP}\label{general2R2CF}

A Sticky PYP is defined by the  following  general properties:

\begin{enumerate}

\item[Property 1:] Set $\mathcal{G}$ comprises a countable number of generative \textit{groups}. Each group $g \in \mathcal{G}$ contains a countable number of group-specific regular PYPs. The PYPs are identified by a combination of the group label $g$ and an integer-valued \textit{state},~$s \in \mathcal{S}$. In other words, the  PYPs have bivariate  labels, $(g,s)\in \mathcal{G}\times \mathcal{S}$.

\item[Property 2:]
For every  $(g,s) \in \mathcal{G}\times \mathcal{S}$, let the corresponding PYP be $\mathcal{W}_{gs}(d_{s}, \alpha_{s}, W_{s})$, of which the $T$-variate base distribution $W_s$, discount parameter $d_s \in [0,1)$, and mass parameter $\alpha_s>0$ depend on the state $s$.  If  set $\mathcal{S}$ contains multiple states, assume that the  base distributions $\{W_{s}: s \in \mathcal{S}\}$ are such that two PYPs associated with unequal  states   almost surely have non-intersecting sets of atoms.

Let  distribution  $\mathcal{P}_{gs}$ be an independent realizations of  the corresponding PYP:
\[
\mathcal{P}_{gs} \,\big|\, d_{s}, \alpha_{s}, W_{s} \stackrel{indep} \sim \mathcal{W}_{gs}(d_{s}, \alpha_{s}, W_{s}), \quad (g,s) \in \mathcal{G}\times \mathcal{S}.
\]
Consequently, for each   $s$,  the distributions $\{\mathcal{P}_{gs}:g \in \mathcal{G}\}$ share the same atoms as  base distribution $W_s$; however,  the probabilities associated with the atoms  depend on  the group-state combination.

\smallskip

\item[Property 3:]
For probe $j=1,\ldots,p$, the label of the PYP  generating the random effect $\boldsymbol{\theta}_j$ is denoted by $(g_j,s_j) \in \mathcal{G}\times \mathcal{S}$, and 
$
 \boldsymbol{\theta}_j \mid \mathcal{P}_{g_j s_j} \stackrel{indep}\sim \mathcal{P}_{g_j s_j}.
 $

\smallskip

\item[Property 4:]\label{SP1} Given the group $g_j$ of the $j^{th}$ probe, the state  has the distribution:
    \begin{equation*}
 s_j \mid g_j \sim \mathcal{Q}_{g_j} 
 \end{equation*}
 where, for every $g \in \mathcal{G}$, $\mathcal{Q}_g$ denotes a group-specific probability mass function on the set $\mathcal{S}$; thus, $\sum_{s \in \mathcal{S}}\mathcal{Q}_{g}(s) = 1$.

\smallskip

\item[Property 5:]
\textit{(Markov property)} \quad 
For the first probe,  group  $g_1$  follows a categorical distribution, $\mathcal{F}_{1}$, on the set $\mathcal{G}$. For the subsequent probes $j>1$,  given the variables associated with the preceding probes, the mass function  of group variable $g_j$ generally depends on  probe index $j$, random vector $\boldsymbol{\theta}_{j-1}$, and inter-probe distance~$e_{j-1}$:
 \begin{equation*}
 g_j  \sim \mathcal{F}_{j\boldsymbol{\theta}_{j-1}e_{j-1}} 
 \end{equation*}
 where,  for every $\boldsymbol{\theta} \in \mathcal{R}^T$ and $e>0$, $\mathcal{F}_{j\boldsymbol{\theta}e}$ denotes a probability mass function on the set $\mathcal{G}$, so that $\sum_{g \in \mathcal{G}}\mathcal{F}_{j\boldsymbol{\theta}e}(g) = 1$.

\end{enumerate}

Some popular examples  are presented in Table \ref{T1} of Supplementary Material. As suggested by the form of  distribution $\mathcal{F}_{j\boldsymbol{\theta}  e}$ in the table, rows 1--5 of Table \ref{T1} display zero order Sticky~PYPs, e.g., hierarchical Pitman-Yor processes (HPYP) \citep{Teh_etal_2006,camerlenghi2019distribution}. Rows 6 and 7 are first order Sticky~PYPs involving  multiple groups and single states. The  mass functions $\mathcal{F}_{j\boldsymbol{\theta}  e}$ do not depend on   inter-probe distance~$e$ in the examples.

For a detailed discussion of a specific mixture model, consider the hierarchical Dirichlet process (HDP) of  \cite{Teh_etal_2006}, in which   $p$ objects are organized into $K$ groups with known group memberships $g_1,\ldots,g_p$. Let global measure $W_{\infty}$ be  a Dirichlet process realization, and therefore, a countably infinite distribution in $\mathcal{R}^T$.
 The random effects $\boldsymbol{\theta}_1,\ldots,\boldsymbol{\theta}_p$  are distributed as:
\begin{align}
 \boldsymbol{\theta}_j \mid G_{g_j} &\stackrel{indep}\sim G_{g_j}, \quad\text{$j=1,\ldots,p,$}\notag\\
 G_{g} \mid \alpha, W_{\infty}&\stackrel{i.i.d.}\sim \text{DP}\bigl(\alpha,W_{\infty}\bigr), \quad\text{$g=1,\ldots,K$.} \label{HDP}
\end{align}
As summarized in the fourth row of Table \ref{T1}, the model can be recast as a generalized Sticky PYP as follows:
\begin{enumerate}
    \item[Property 1:] The set of group labels, $\mathcal{G}=\{1,\ldots,K\}$.  An HDP is a single-state model  with state set  $\mathcal{S}=\{1\}$, i.e., $s_j=1$ for all $p$ objects. To simplify the notation, we henceforth drop the subscript $s$.
    
    \item[Property 2:]  For group $g=1,\ldots,K$, the  associated  PYP is  $\mathcal{W}_{g}\bigl(0,\alpha,W_{\infty}\bigr)=$ $\text{DP}\bigl(\alpha,W_{\infty}\bigr)$.
    
     \item[Property 3:]  
Since distribution $\mathcal{P}_{g}$ in the generalized Sticky PYP formulation is a random realization of   $\mathcal{W_{\infty}}_{g}\bigl(0,\alpha,W_{\infty}\bigr)=\text{DP}\bigl(\alpha,W_{\infty}\bigr)$, we identify   $\mathcal{P}_{g}=G_g$ in expression~(\ref{HDP}). 
     
     \item[Property 4:] An HDP is a single-state model; therefore, trivially, $\mathcal{Q}_{g}=1_{\{1\}}$.
     
      \item[Property 5:] Group variables $g_1,\ldots,g_p$ are known. Hence, $\mathcal{F}_{1}=1_{\{g_1\}}$  and $\mathcal{F}_{j\boldsymbol{\theta},e}=1_{\{g_j\}}$,  $j=2,\ldots,p$.
\end{enumerate}

The 2R2CF  used in differential methylation is a Sticky PYP with  two groups and two states.
 A key characteristic distinguishing   it    from the other first  order models, and indeed, all the models listed in Table \ref{T1}, is that the 2R2CF makes a probabilistic, rather than deterministic, assumption for group variable $g_j$ that varies with the inter-probe distances.
 As noted in Section \ref{Sticky_PYPs_DA}, the 2R2CF behaves similarly to two-state hidden Markov models for very small inter-probe distances and similarly to finite mixture models for relatively large distances. For differential analysis, this offers a key advantage in  datasets with widely varying inter-probe distances by allowing the differential and non-differential probes  to have  different  cluster allocation patterns depending on the differential statuses and distances of  adjacent probes. These  interpretable features of the 2R2CF process are  facilitated by the  two-group, two-state~construct.

{
\begin{table}
\footnotesize
\begin{center}
\renewcommand{\arraystretch}{1.5}
\begin{tabular}{l | l l l  }
 \hline\hline
 \textbf{Model} &$\mathcal{G}$   &$\mathcal{W}_{gs}(d_{s}, \alpha_{s}, W_{s})$   &$\mathcal{F}_{j\boldsymbol{\theta}e}$  \\ 
 \hline
  \hline
  Finite mixture &$\{1\}$   &$\mathcal{W}(0, \alpha, W_K)$,      &$1_{\{1\}}$  
 \\ \hline
  Dirichlet process &$\{1\}$    &$\mathcal{W}(0, \alpha, W)$   &$1_{\{1\}}$ 
  \\
  \hline
  PYP &$\{1\}$   &$\mathcal{W}(d, \alpha, W)$   &$1_{\{1\}}$ 
  \\
  \hline
  HDP &$\mathcal{N}_K$       &$\mathcal{W}_{g}(0, \alpha, W_{\infty})$   &$1_{\{g_j\}}$; known $g_j \in \mathcal{G}$ 
  \\\hline
  HPYP &$\mathcal{N}_K$       &$\mathcal{W}_{g}(d, \alpha, W_{\infty})$   &$1_{\{g_j\}}$; known $g_j \in \mathcal{G}$ 
  \\
  \hline
  Finite HMM &$\mathcal{N}_K$  &$\mathcal{W}_g(0, \alpha, W_K)$      &Point mass at $\sum_{v=1}^K v\cdot \mathcal{I}(\boldsymbol{\theta}=\boldsymbol{\phi}_{v})$
  \\\hline
  HDP-HMM &$\mathcal{N}$   &$\mathcal{W}_{g}(0, \alpha, W_{\infty})$  &Point mass at $\sum_{v=1}^\infty v\cdot \mathcal{I}(\boldsymbol{\theta}=\boldsymbol{\phi}_{v})$ 
  \\
  \hline
\hline
\end{tabular}
\end{center}
\caption{Examples of Sticky PYPs. The  examples listed here correspond to a singleton state set, $\mathcal{S}=\{1\}$,  and degenerate distribution, $\mathcal{Q}_{g}=1_{\{1\}}$.  Set $\mathcal{N}$ represents the natural numbers, $\mathcal{N}_K=$ $\{1,\ldots,K\}$ for some $K>1$,  $d \in [0,1)$, and   $W$ is an arbitrary distribution in $\mathcal{R}^T$.  Distribution $W_{K}$ is finite with $T$-variate atoms $\{\boldsymbol{\phi}_{v}\}_{v=1}^{K}$. Distribution $W_{\infty}$ is countably infinite with $T$-variate atoms $\{\boldsymbol{\phi}_{v}\}_{v=1}^{\infty}$. In an HDP or HDP-HMM, $W_{\infty}$ represents  a Dirichlet process realization. For an HPYP, $W_{\infty}$  represents  a PYP  realization. Refer to   Section~\ref{general2R2CF} for the notation and further discussion. }\label{T1}
\end{table}
}

\begin{table}
	\centering
	\footnotesize
\begin{tabular}{c c | c c c c}
 & & \multicolumn{2}{c}{Low noise}  &\multicolumn{2}{c}{High noise}   \\
 \hline
 & & High correlation  & No correlation  & High correlation & No correlation   \\  \hline
\multirow{7}{0.1\textwidth}{\centering AUC} & BayesDiff & \textbf{0.995} & \textbf{0.970} & \textbf{0.944} & 0.830 \\
 & ANOVA & 0.964 & 0.949 & 0.888 & \textbf{0.864} \\
 & Kruskal-Wallis & 0.958 & 0.944 & 0.878 & 0.854 \\
 & COHCAP & 0.955 & 0.941 & 0.861 & 0.841 \\
 & Methylkit & 0.955 & 0.942 & 0.869 & 0.848 \\
 & BiSeq & 0.949 & 0.933 & 0.867 & 0.845 \\
 & RADMeth & 0.959 & 0.944 & 0.876 & 0.852 \\ \hline
 \multirow{7}{0.1\textwidth}{\centering $\textrm{AUC}_{20}$} & BayesDiff & \textbf{0.988} & \textbf{0.926} & \textbf{0.884} & \textbf{0.636} \\
 & ANOVA & 0.891 & 0.851 & 0.684 & 0.628 \\
 & Kruskal-Wallis & 0.876 & 0.833 & 0.657 & 0.597 \\
 & COHCAP & 0.856 & 0.820 & 0.602 & 0.556 \\
 & Methylkit & 0.858 & 0.819 & 0.633 & 0.579 \\
 & BiSeq & 0.822 & 0.781 & 0.607 & 0.559 \\
 & RADMeth & 0.871 & 0.833 & 0.639 & 0.588 \\ \hline
 \multirow{7}{0.1\textwidth}{\centering $\textrm{AUC}_{10}$} & BayesDiff & \textbf{0.985} & \textbf{0.901} & \textbf{0.857} & \textbf{0.565} \\
 & ANOVA & 0.849 & 0.804 & 0.586 & 0.529 \\
 & Kruskal-Wallis & 0.827 & 0.773 & 0.551 & 0.493 \\
 & COHCAP & 0.797 & 0.751 & 0.492 & 0.433 \\
 & Methylkit & 0.794 & 0.744 & 0.523 & 0.458 \\
 & BiSeq & 0.726 & 0.673 & 0.482 & 0.424 \\
 & RADMeth & 0.817 & 0.775 & 0.531 & 0.471 \\ 
 \end{tabular}
\caption{Areas under ROC curves for the different methods (rows) under the four simulation scenarios (columns). See the text in the paper for further discussion.}
\label{table_sim1}
\end{table}

\begin{figure}[h]
\centering
\includegraphics[width=0.8\textwidth]{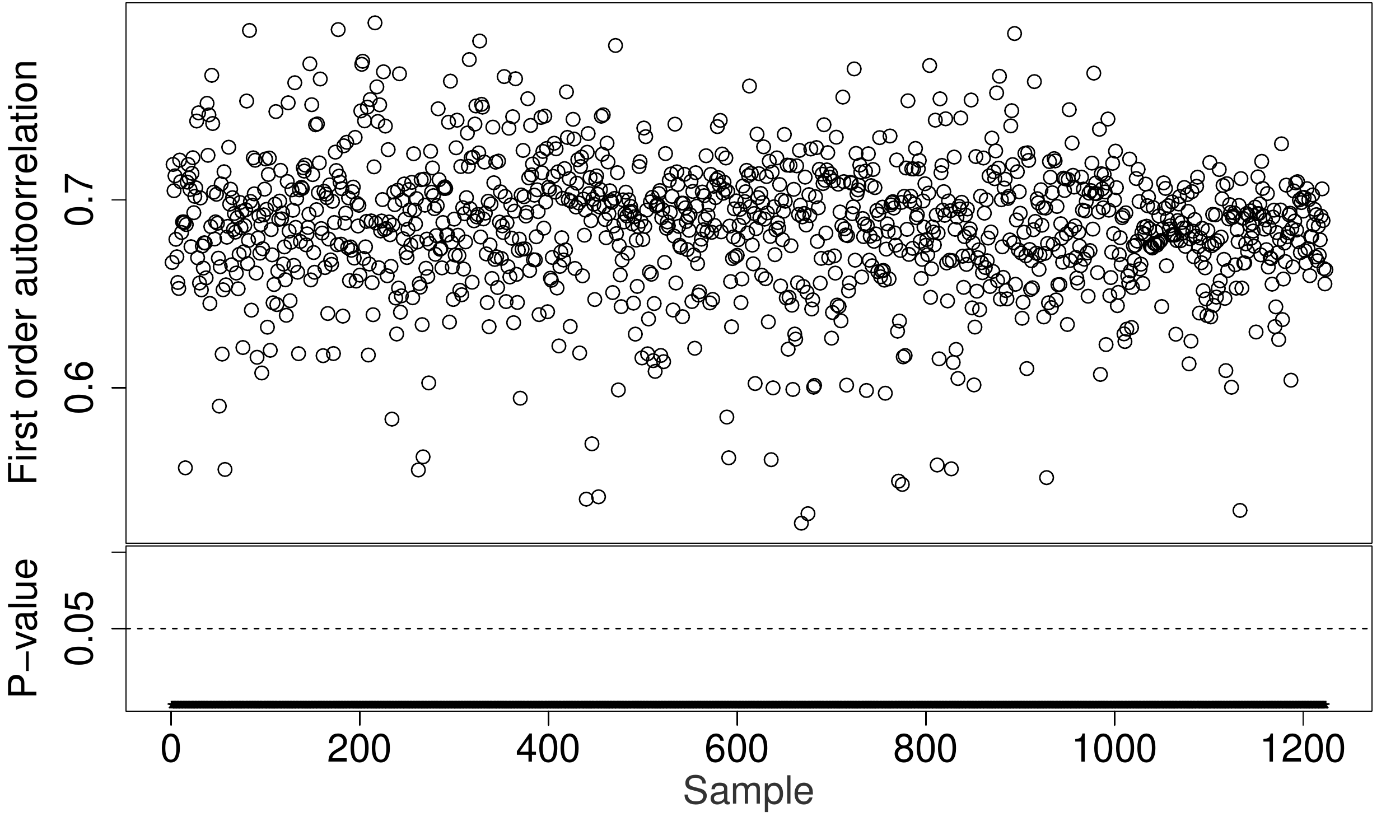}
	\caption{Exploratory analysis of DNA methylation profiles of upper GI cancer samples  from TCGA. See the text in the  paper for further explanation. }
	\label{intro_data_plots2}
\end{figure}


\begin{figure}
	\centering
    \captionsetup[subfigure]{justification=centering}
\begin{subfigure}{0.47\textwidth}
\centering
\includegraphics[width=\textwidth]{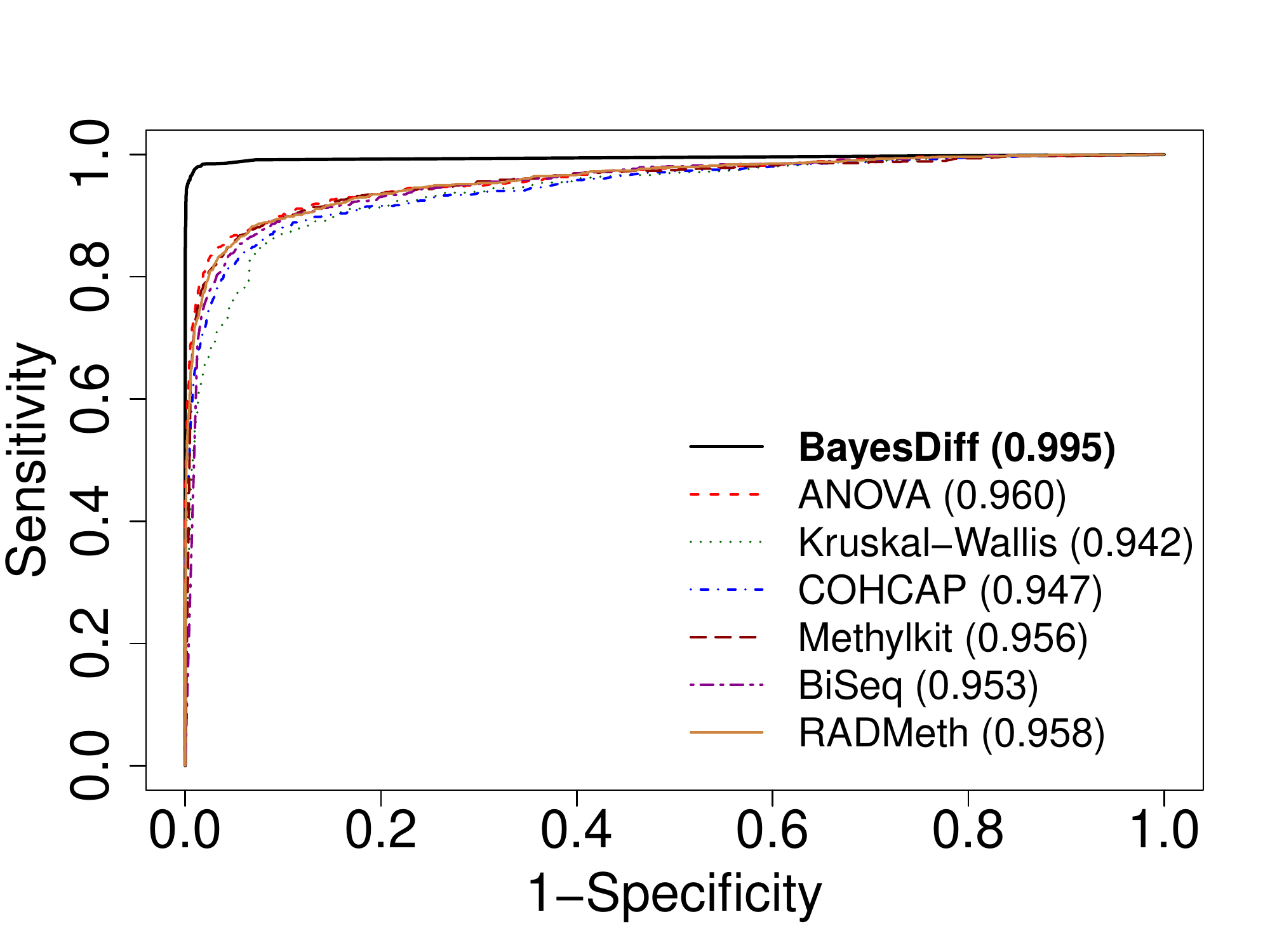}
\caption{Low noise, high correlation}
\end{subfigure}%
\quad
\begin{subfigure}{0.47\textwidth}
\centering
\includegraphics[width=\textwidth]{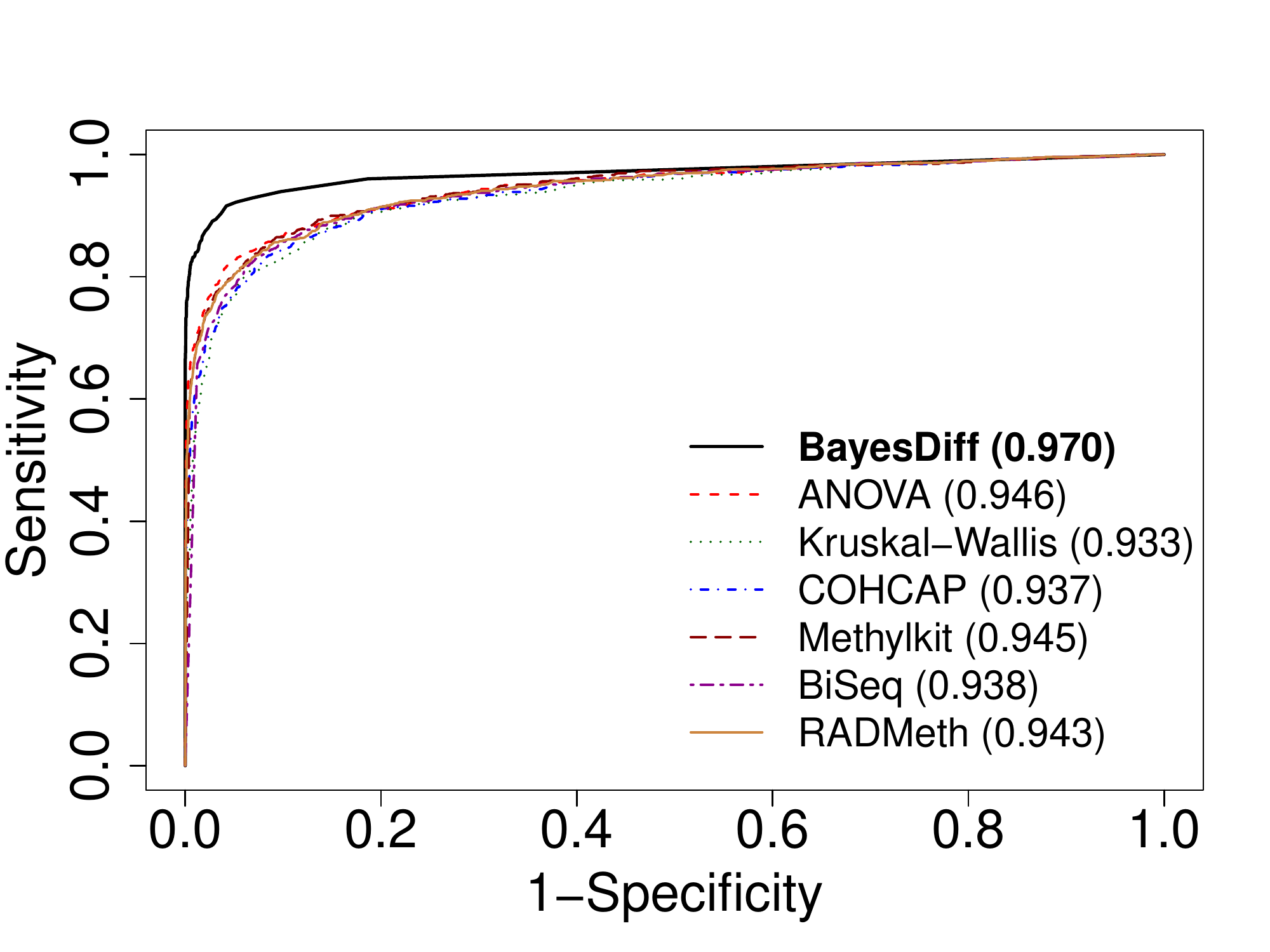}
\caption{Low noise, no correlation}
\end{subfigure}
\begin{subfigure}{0.47\textwidth}
\centering
\includegraphics[width=\textwidth]{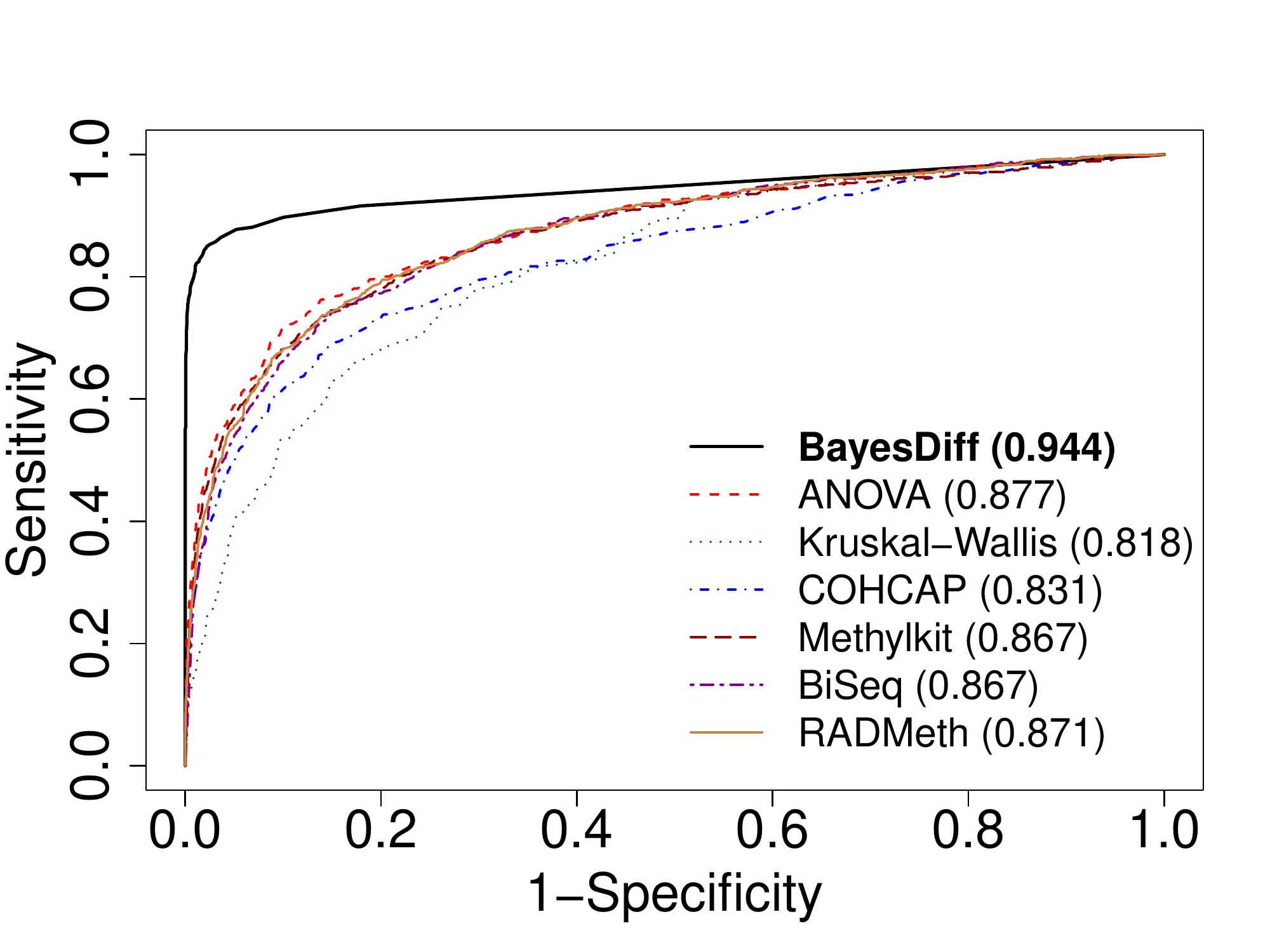}
\caption{High noise, high correlation}
\end{subfigure}%
\quad
\begin{subfigure}{0.47\textwidth}
\centering
\includegraphics[width=\textwidth]{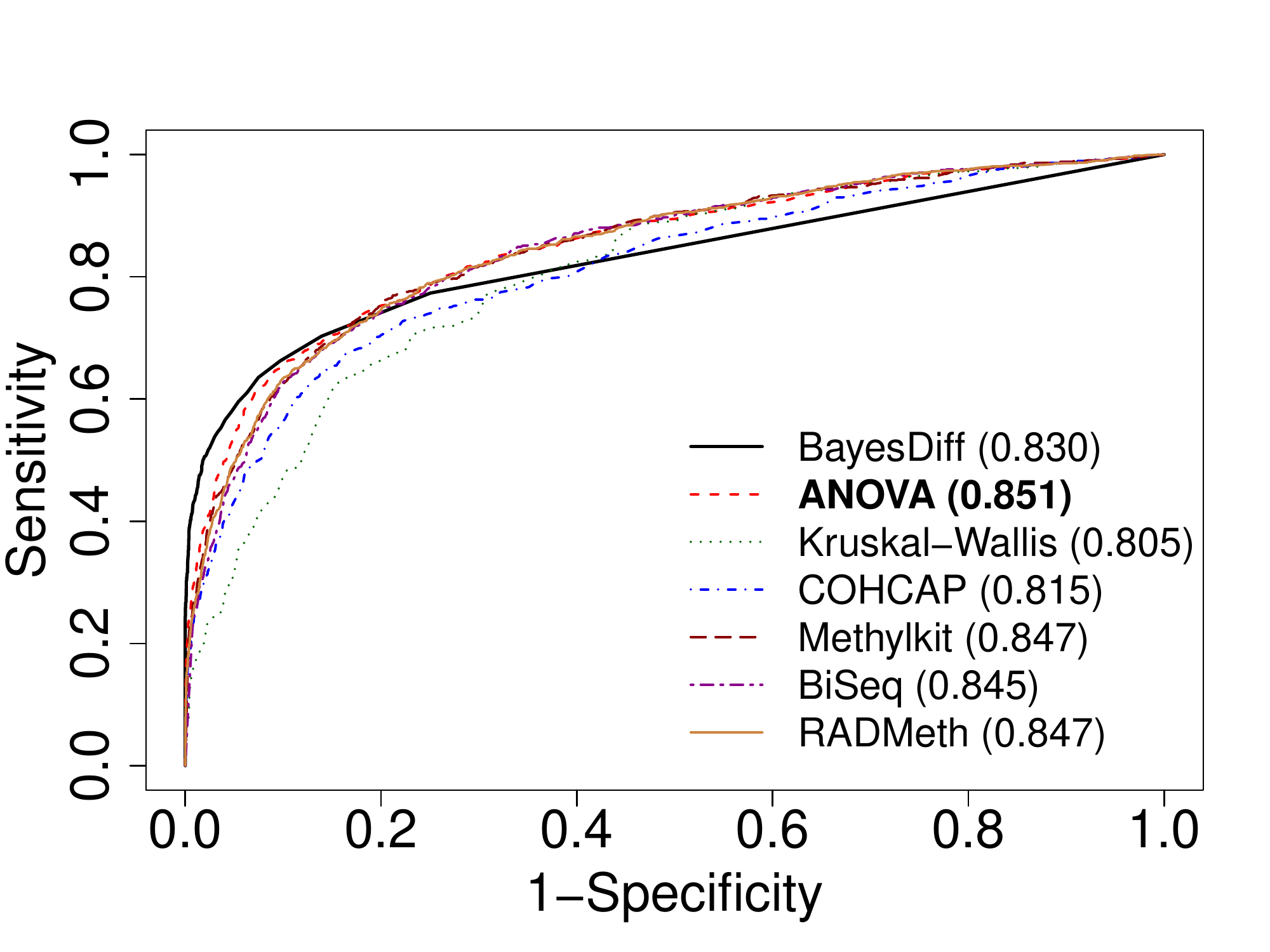}
\caption{High noise, no correlation}
\end{subfigure}
	\caption{ROC curves, averaged over 20 simulated datasets, for the seven methods and the four simulation scenarios. The numbers in  parentheses represent AUCs.}
	\label{sim1_ROC_avg}
\end{figure}


\begin{figure}
	\centering
    \captionsetup[subfigure]{justification=centering}
\begin{subfigure}{0.6\textwidth}
\centering
\includegraphics[width=\textwidth]{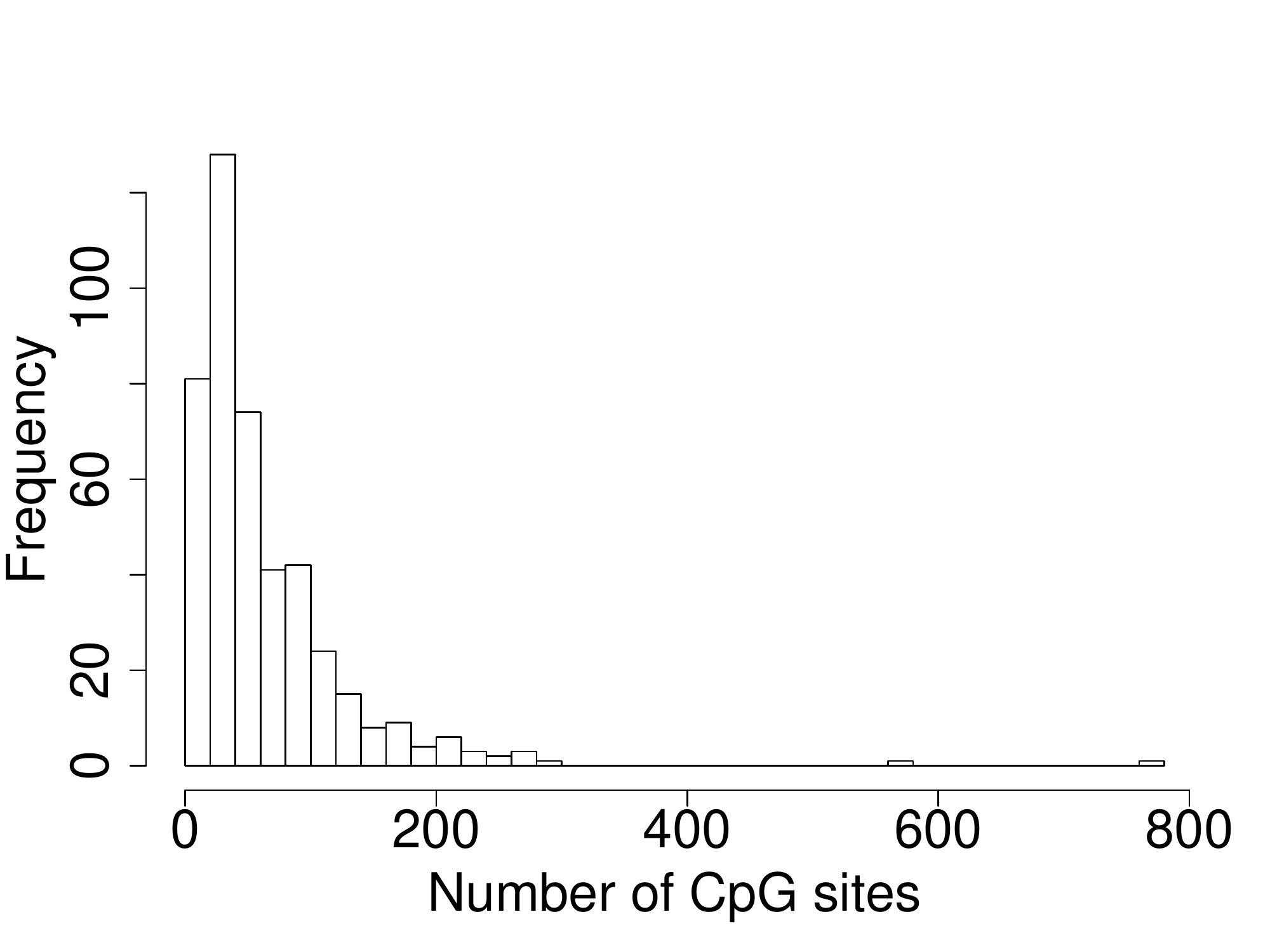}
\caption{Histogram of the number of gene-specific CpG~sites}
\end{subfigure}%
\quad
\begin{subfigure}{0.6\textwidth}
\centering
\includegraphics[width=\textwidth]{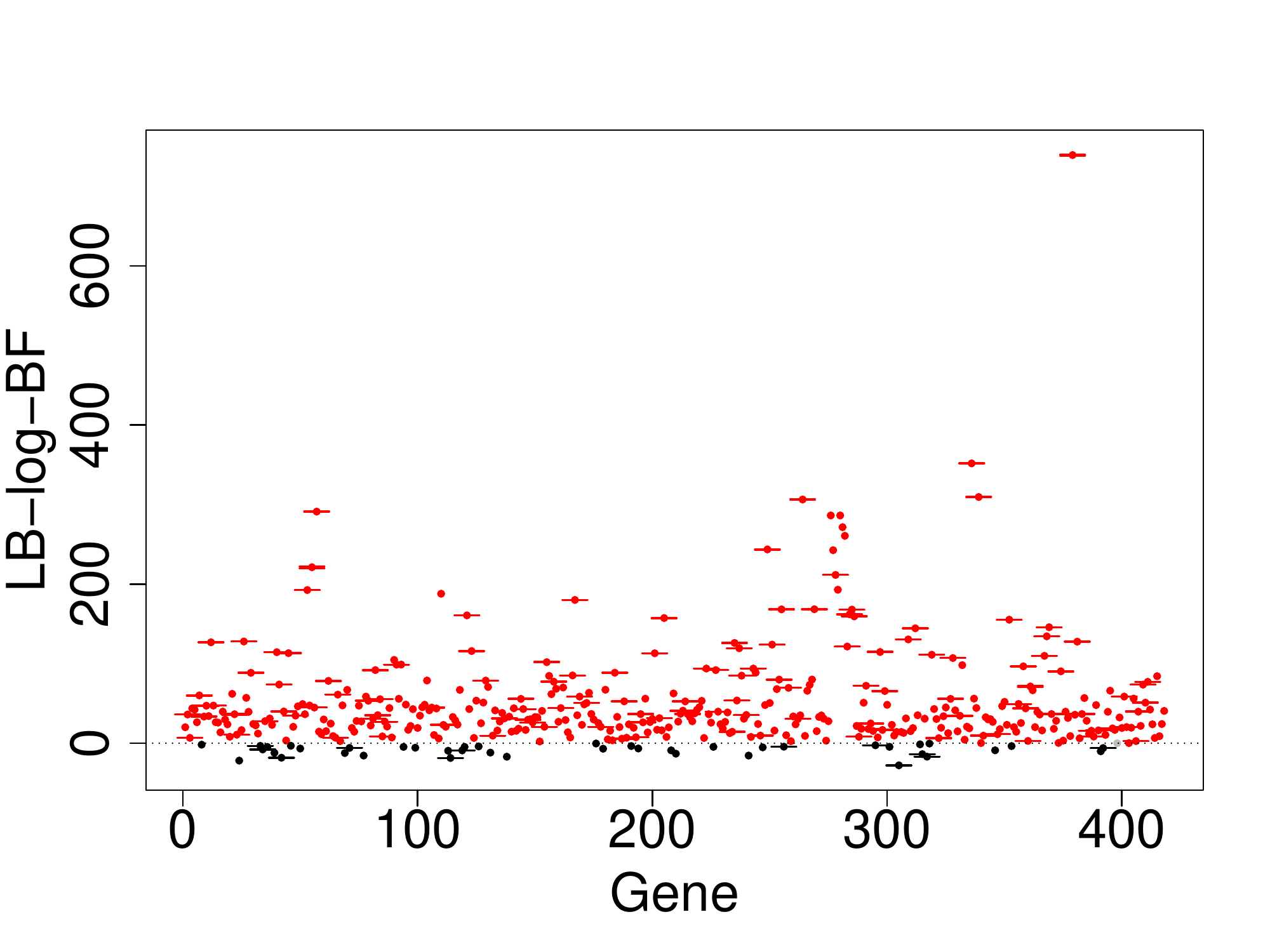}
\caption{95\% credible intervals for  lower bounds of log-Bayes factors of first order versus zero-order models. The intervals whose lower limits are positive are marked in red.}
\end{subfigure}
	\caption{Data analysis plots. See the Section \ref{data_analysis} text for further explanation.}
	\label{appl_plots}
\end{figure}

	



\begin{table}
\centering
\begin{tabular}{ccc|ccc}
  \hline \hline
  \multicolumn{6}{c}{\large \textbf{BayesDiff}}\TBstrut\\ 
  \hline
\multirow{2}{*}{Gene} & DM & Largest & \multirow{2}{*}{Gene} & DM & Largest \\
 & Proportion & Difference & & Proportion & Difference\\
  \hline
EDNRB & 1.00 & STAD$\uparrow$  LIHC$\downarrow$ & SLITRK1 & 0.96 & STAD$\uparrow$  LIHC$\downarrow$ \\ 
  PCLO & 1.00 & PAAD$\uparrow$  LIHC$\downarrow$ & FLRT2 & 0.96 & STAD$\uparrow$  LIHC$\downarrow$ \\ 
  PREX2 & 1.00 & STAD$\uparrow$  LIHC$\downarrow$ & KCNA1 & 0.96 & STAD$\uparrow$  LIHC$\downarrow$ \\ 
  SLIT2 & 1.00 & STAD$\uparrow$  LIHC$\downarrow$ & TRPA1 & 0.96 & STAD$\uparrow$  LIHC$\downarrow$ \\ 
  SLITRK2 & 1.00 & STAD$\uparrow$  LIHC$\downarrow$ & ADCY8 & 0.96 & STAD$\uparrow$  LIHC$\downarrow$ \\ 
  SORCS3 & 1.00 & STAD$\uparrow$  LIHC$\downarrow$ & CTNNA2 & 0.95 & PAAD$\uparrow$  LIHC$\downarrow$ \\ 
  SPHKAP & 1.00 & ESCA$\uparrow$  LIHC$\downarrow$ & GRIA2 & 0.95 & STAD$\uparrow$  LIHC$\downarrow$ \\ 
  SPTA1 & 1.00 & PAAD$\uparrow$  LIHC$\downarrow$ & ADGRL3 & 0.94 & PAAD$\uparrow$  LIHC$\downarrow$ \\ 
  UNC13C & 1.00 & PAAD$\uparrow$  LIHC$\downarrow$ & LRRC7 & 0.94 & STAD$\uparrow$  LIHC$\downarrow$ \\ 
  XIRP2 & 1.00 & PAAD$\uparrow$  LIHC$\downarrow$ & ERBB4 & 0.93 & PAAD$\uparrow$  LIHC$\downarrow$ \\ 
  ZNF804B & 1.00 & STAD$\uparrow$  LIHC$\downarrow$ & PCDH10 & 0.93 & STAD$\uparrow$  LIHC$\downarrow$ \\
  TSHZ3 & 0.97 & STAD$\uparrow$  LIHC$\downarrow$ & SOX11 & 0.93 & STAD$\uparrow$  LIHC$\downarrow$ \\ 
  MYO3A & 0.97 & STAD$\uparrow$  LIHC$\downarrow$ & NLGN4X & 0.93 & PAAD$\uparrow$  LIHC$\downarrow$ \\ 
  ABCC9 & 0.97 & ESCA$\uparrow$  LIHC$\downarrow$ & NBEA & 0.93 & PAAD$\uparrow$  LIHC$\downarrow$ \\ 
  EPB41L3 & 0.97 & STAD$\uparrow$  LIHC$\downarrow$ & CNTN1 & 0.92 & STAD$\uparrow$  LIHC$\downarrow$ \\ 
  FBN2 & 0.97 & STAD$\uparrow$  LIHC$\downarrow$ & GRM5 & 0.92 & PAAD$\uparrow$  LIHC$\downarrow$ \\ 
  PCDH17 & 0.96 & STAD$\uparrow$  LIHC$\downarrow$ & PTPRZ1 & 0.91 & STAD$\uparrow$  PAAD$\downarrow$ \\ 
  CDH8 & 0.96 & STAD$\uparrow$  LIHC$\downarrow$ & EPHA5 & 0.91 & STAD$\uparrow$  LIHC$\downarrow$ \\ 
   \hline \hline
\end{tabular}
    \caption{For \textbf{BayesDiff}, genes with the overall proportion of differentially methylated CpG sites exceeding 0.9. The ``Largest Difference'' column displays which pair-wise difference between the four cancer types is the largest, with the symbol ``$\uparrow$ ($\downarrow$)'' indicating higher (lower) methylation level for one cancer type relative to the other.}
  \label{table_list_gene_prop}%
\end{table}

	\begin{table}
		\centering
		\begin{tabular}{lc|lc}
			\hline \hline
			\multicolumn{4}{c}{\large \textbf{BayesDiff}}\TBstrut\\ 
  \hline
			Gene & Number of probes & Gene & Number of probes \\ 
			\hline
			ABCC9 &  36 & NLGN4X &  28 \\ 
			ADCY8 &  24 & PCDH10 &  30 \\ 
			ADGRL3 &  35 & PCDH17 &  27 \\ 
			CDH8 &  26 & PCLO &  20 \\ 
			CNTN1 &  25 & PREX2 &  21 \\ 
			CTNNA2 &  85 & PTPRZ1 &  23 \\ 
			EDNRB &  56 & SLIT2 &  29 \\ 
			EPB41L3 &  34 & SLITRK1 &  26 \\ 
			EPHA5 &  22 & SLITRK2 &  28 \\ 
			ERBB4 &  30 & SORCS3 &  39 \\ 
			FBN2 &  30 & SOX11 &  44 \\ 
			FLRT2 &  51 & SPHKAP &  17 \\ 
			GRIA2 &  20 & SPTA1 &  16 \\ 
			GRM5 &  25 & TRPA1 &  25 \\ 
			KCNA1 &  25 & TSHZ3 &  39 \\ 
			LRRC7 &  34 & UNC13C &  11 \\ 
			MYO3A &  38 & XIRP2 &  19 \\ 
			NBEA &  55 & ZNF804B &  21 \\ 
			\hline \hline
		\end{tabular}
		\caption{Number of included CpG sites for the top methylated genes listed in  Table \ref{table_list_gene_prop} for BayesDiff.} 
		\label{table_list_num_probe}%
	\end{table}

\begin{table}[htbp]
  \centering
    \begin{tabular}{ccccc}
  \hline \hline
  \multicolumn{5}{c}{\large \textbf{BayesDiff and ANOVA}}\TBstrut\\ 
  \hline
\multirow{2}{*}{Gene} & DM Proportion & Largest Difference & DM Proportion & Largest Difference \\
 & BayesDiff & BayesDiff & ANOVA & ANOVA \\
  \hline 
    PCLO  & 1.00  & PAAD↑ LIHC↓ & 1.00  & PAAD↑ LIHC↓ \\
    UNC13C & 1.00  & PAAD↑ LIHC↓ & 1.00  & PAAD↑ ESCA↓ \\
    XIRP2 & 1.00  & PAAD↑ LIHC↓ & 1.00  & PAAD↑ STAD↓ \\
    ZNF804B & 1.00  & STAD↑ LIHC↓ & 1.00  & ESCA↑ PAAD↓ \\
    FBN2  & 0.97  & STAD↑ LIHC↓ & 0.93  & PAAD↑ STAD↓ \\
    FLRT2 & 0.96  & STAD↑ LIHC↓ & 0.92  & STAD↑ PAAD↓ \\
    ERBB4 & 0.93  & PAAD↑ LIHC↓ & 0.93  & STAD↑ PAAD↓ \\
    PTPRZ1 & 0.91  & STAD↑ PAAD↓ & 0.91  & PAAD↑ LIHC↓ \\
      \hline \hline
    \end{tabular}%
  \caption{For both \textbf{BayesDiff and ANOVA} methods, the common set of genes detected with the overall proportion of differentially methylated CpG sites exceeding 0.9. For each method, the ``Largest Difference'' column displays which pair-wise difference between the four cancer types is the largest, with the symbol ``$\uparrow$ ($\downarrow$)'' indicating higher (lower) methylation level for one cancer type relative to the other.}
  \label{table_list_overlap_gene_prop}%
\end{table}%

\begin{table}[htbp]
  \centering
    \begin{tabular}{ccc|ccc}
\hline \hline
\multicolumn{6}{c}{\large \textbf{ANOVA, not BayesDiff}}\TBstrut\\ 
  \hline
\multirow{2}{*}{Gene} & DM & Largest & \multirow{2}{*}{Gene} & DM & Largest \\
 & Proportion & Difference & & Proportion & Difference\\
 \hline
    LRP12 & 1.00  & LIHC↑ STAD↓ & EPHA6 & 0.95  & STAD↑ PAAD↓ \\
    PTPRM & 1.00  & STAD↑ LIHC↓ & RELN  & 0.95  & STAD↑ LIHC↓ \\
    SALL1 & 0.96  & STAD↑ LIHC↓ & COL6A6 & 0.94  & STAD↑ LIHC↓ \\
    TIAM1 & 0.96  & LIHC↑ PAAD↓ & MUC16 & 0.94  & PAAD↑ LIHC↓ \\
    MXRA5 & 0.96  & LIHC↑ PAAD↓ & HCN1  & 0.91  & PAAD↑ ESCA↓ \\
    ZFHX4 & 0.96  & LIHC↑ STAD↓ & RIMS2 & 0.91  & STAD↑ LIHC↓ \\
    \hline \hline
    \end{tabular}%
  \caption{For  \textbf{only ANOVA but not BayesDiff}, genes detected with the overall proportion of differentially methylated CpG sites exceeding 0.9. The ``Largest Difference'' column displays which pair-wise difference between the four cancer types is the largest, with the symbol ``$\uparrow$ ($\downarrow$)'' indicating higher (lower) methylation level for one cancer type relative to the other.}
  \label{table_list_ANOVA_only_gene_prop}%
\end{table}%

\begin{table}[htbp]
  \centering
    \begin{tabular}{ccc|ccc}
\hline \hline
\multicolumn{6}{c}{\large \textbf{BayesDiff, not ANOVA}}\TBstrut\\ 
  \hline
\multirow{2}{*}{Gene} & DM & Largest & \multirow{2}{*}{Gene} & DM & Largest \\
 & Proportion & Difference & & Proportion & Difference\\
 \hline
    EDNRB & 1.00  & STAD↑ LIHC↓ & KCNA1 & 0.96  & STAD↑ LIHC↓ \\
    PREX2 & 1.00  & STAD↑ LIHC↓ & TRPA1 & 0.96  & STAD↑ LIHC↓ \\
    SLIT2 & 1.00  & STAD↑ LIHC↓ & ADCY8 & 0.96  & STAD↑ LIHC↓ \\
    SLITRK2 & 1.00  & STAD↑ LIHC↓ & CTNNA2 & 0.95  & PAAD↑ LIHC↓ \\
    SORCS3 & 1.00  & STAD↑ LIHC↓ & GRIA2 & 0.95  & STAD↑ LIHC↓ \\
    SPHKAP & 1.00  & ESCA↑ LIHC↓ & ADGRL3 & 0.94  & PAAD↑ LIHC↓ \\
    SPTA1 & 1.00  & PAAD↑ LIHC↓ & LRRC7 & 0.94  & STAD↑ LIHC↓ \\
    TSHZ3 & 0.97  & STAD↑ LIHC↓ & PCDH10 & 0.93  & STAD↑ LIHC↓ \\
    MYO3A & 0.97  & STAD↑ LIHC↓ & SOX11 & 0.93  & STAD↑ LIHC↓ \\
    ABCC9 & 0.97  & ESCA↑ LIHC↓ & NLGN4X & 0.93  & PAAD↑ LIHC↓ \\
    EPB41L3 & 0.97  & STAD↑ LIHC↓ & NBEA  & 0.93  & PAAD↑ LIHC↓ \\
    PCDH17 & 0.96  & STAD↑ LIHC↓ & CNTN1 & 0.92  & STAD↑ LIHC↓ \\
    CDH8  & 0.96  & STAD↑ LIHC↓ & GRM5  & 0.92  & PAAD↑ LIHC↓ \\
    SLITRK1 & 0.96  & STAD↑ LIHC↓ & EPHA5 & 0.91  & STAD↑ LIHC↓ \\
    \hline \hline
    \end{tabular}%
  \caption{For  \textbf{only BayesDiff  but not ANOVA}, genes detected with  overall proportion of differentially methylated CpG sites exceeding 0.9. The ``Largest Difference'' column displays which pair-wise difference between the four cancer types is the largest, with the symbol ``$\uparrow$ ($\downarrow$)'' indicating higher (lower) methylation level for one cancer type relative to the other.}
  \label{table_list_BayesDiff_only_gene_prop}%
\end{table}%

\begin{figure}
\centering
\captionsetup[subfigure]{justification=centering}
\begin{subfigure}{\textwidth}
\centering
\includegraphics[trim={0 12pt 0 0},clip,width=0.75\textwidth]{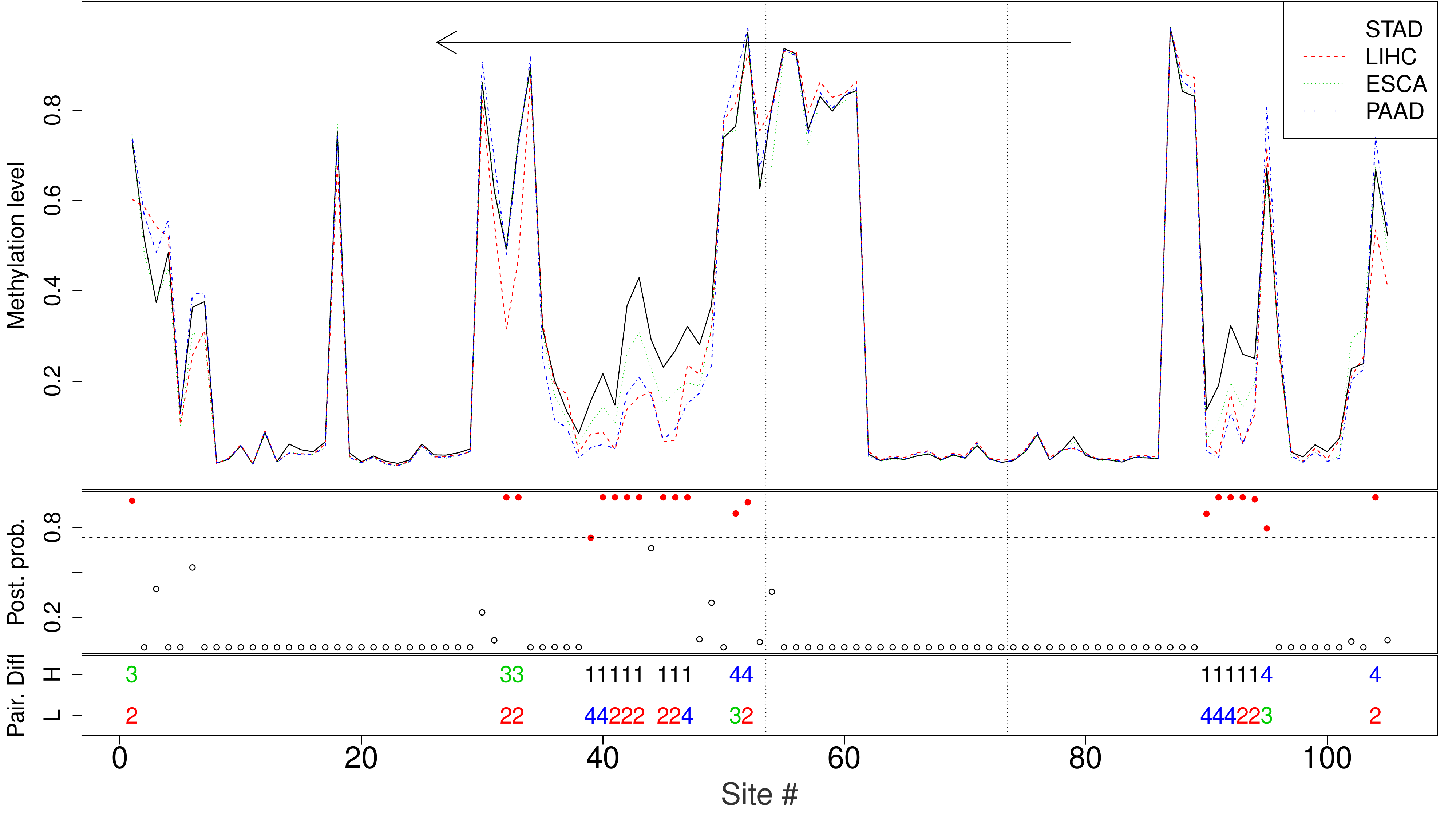}
\caption{Gene TP53}
\vspace{10pt}
\end{subfigure}
\begin{subfigure}{\textwidth}
\centering
\includegraphics[trim={0 12pt 0 0},clip,width=0.75\textwidth]{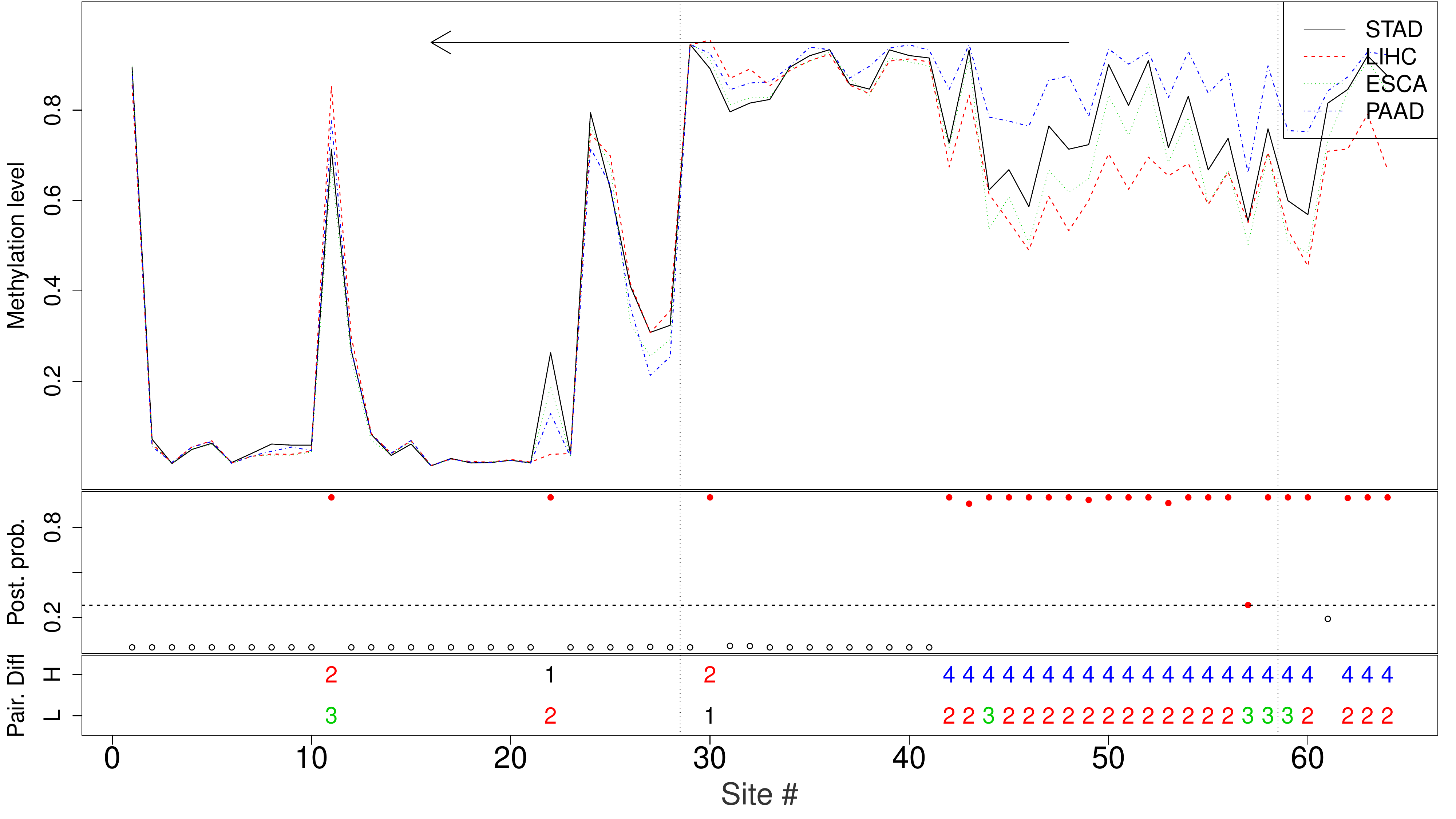}
\caption{Gene TTN}
\vspace{10pt}
\end{subfigure}
	\caption{Detailed differential methylation  results for the top 2 mutated genes. For each gene, the upper panel shows the mean methylation levels. The middle panel shows the posterior probabilities of each CpG site being differentially methylated, with  solid points representing differential methylation and dashed line denoting the corresponding cutoff 
    value. The lower panel indicates the largest pairwise difference between the 4 cancer types. Symbols 1--4 in the lower panel represent  GI cancer types STAD, LIHC, ESCA and PAAD, respectively. 
The vertical dotted lines represent the gene boundaries. The arrow at the top indicates the transcription direction.}
	\label{DA_detail_2genes}
\end{figure}

\begin{figure}
	\centering
    \captionsetup[subfigure]{justification=centering}
\begin{subfigure}{0.48\textwidth}
\centering
\includegraphics[width=\textwidth]{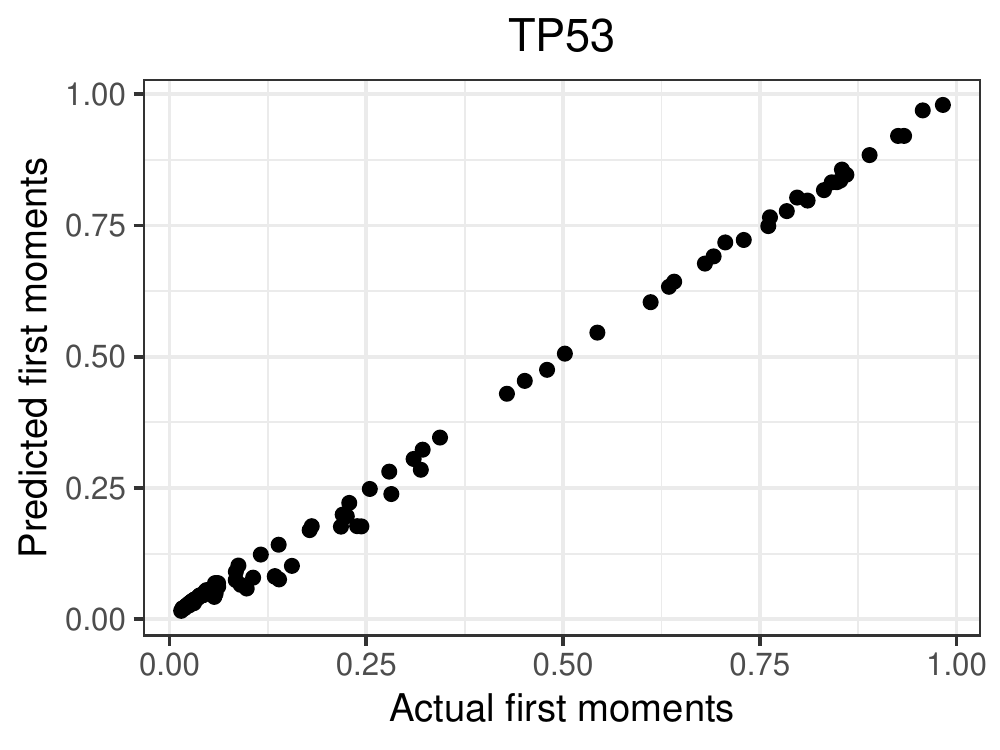}
\caption{First moments for gene TP53}
\end{subfigure}%
\quad
\begin{subfigure}{0.48\textwidth}
\centering
\includegraphics[width=\textwidth]{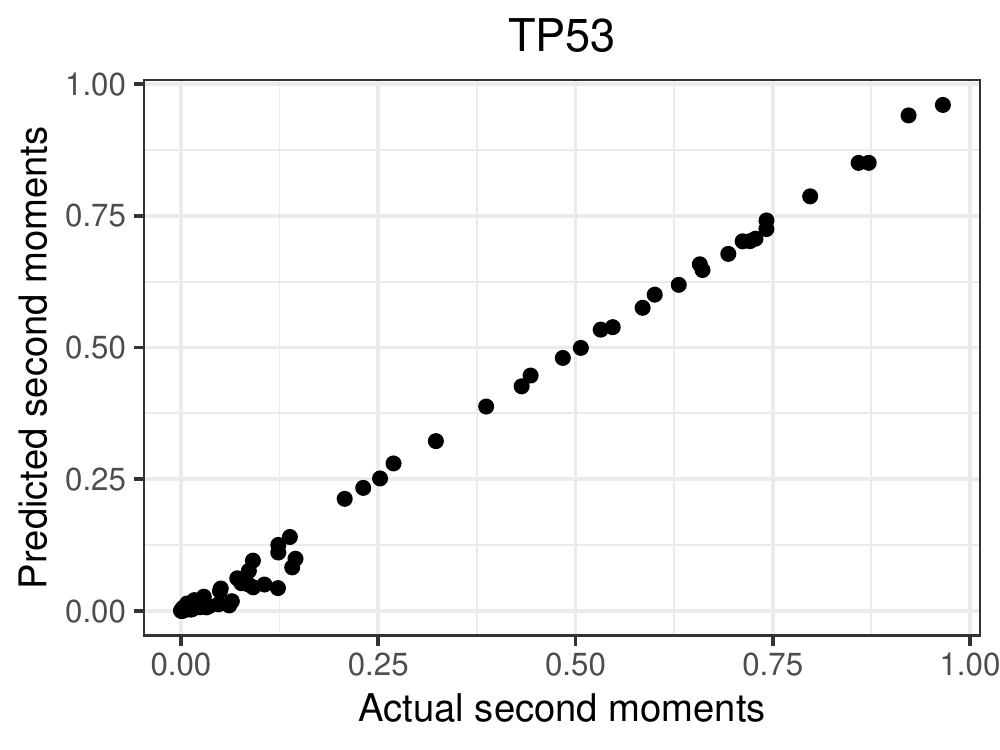}
\caption{Second moments for gene TP53}
\end{subfigure}
\begin{subfigure}{0.48\textwidth}
\centering
\includegraphics[width=\textwidth]{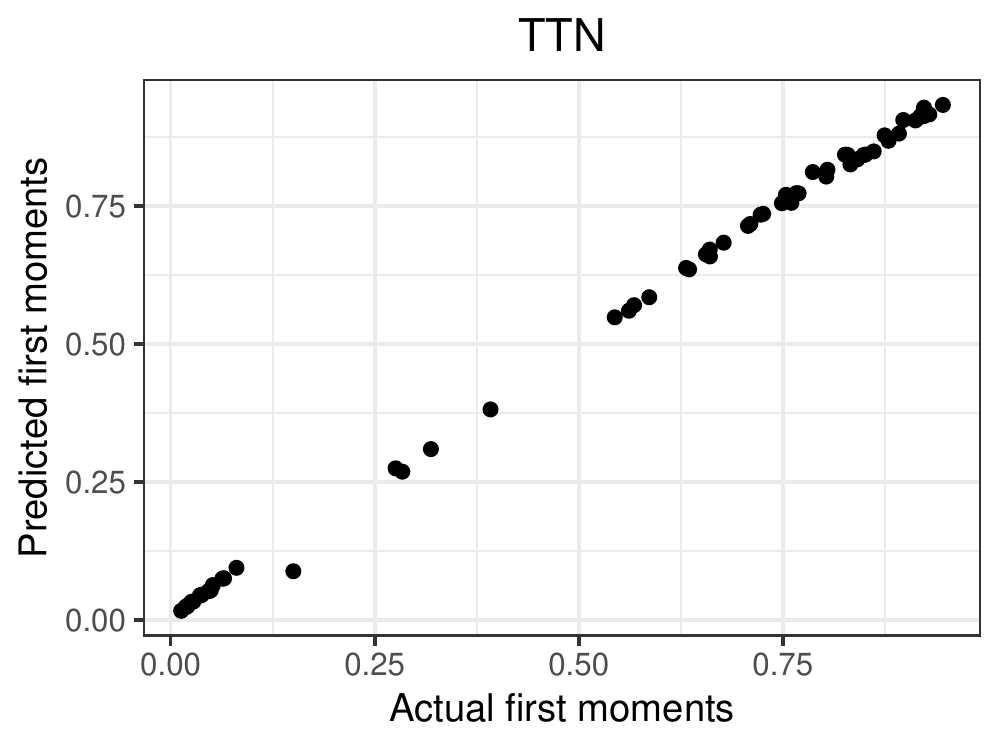}
\caption{First moments for gene TTN}
\end{subfigure}%
\quad
\begin{subfigure}{0.48\textwidth}
\centering
\includegraphics[width=\textwidth]{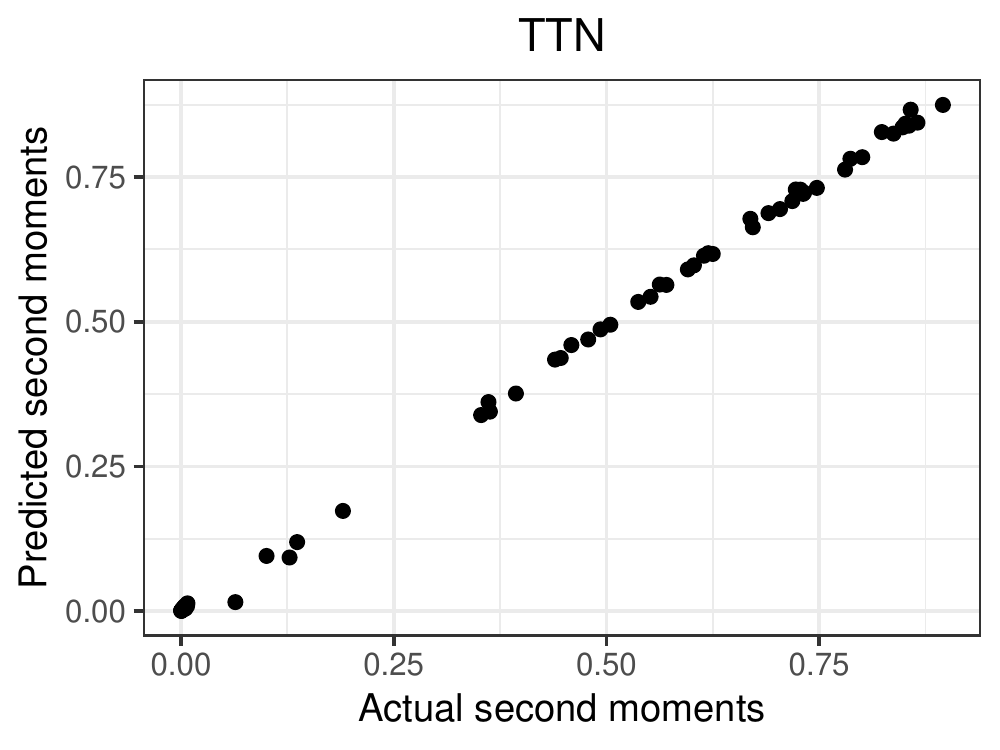}
\caption{Second moments for gene TTN}
\end{subfigure}
	\caption{Comparison of predicted and actual sample moments for the 2 top mutated genes.}
	\label{mean-mean_var-var}
\end{figure}

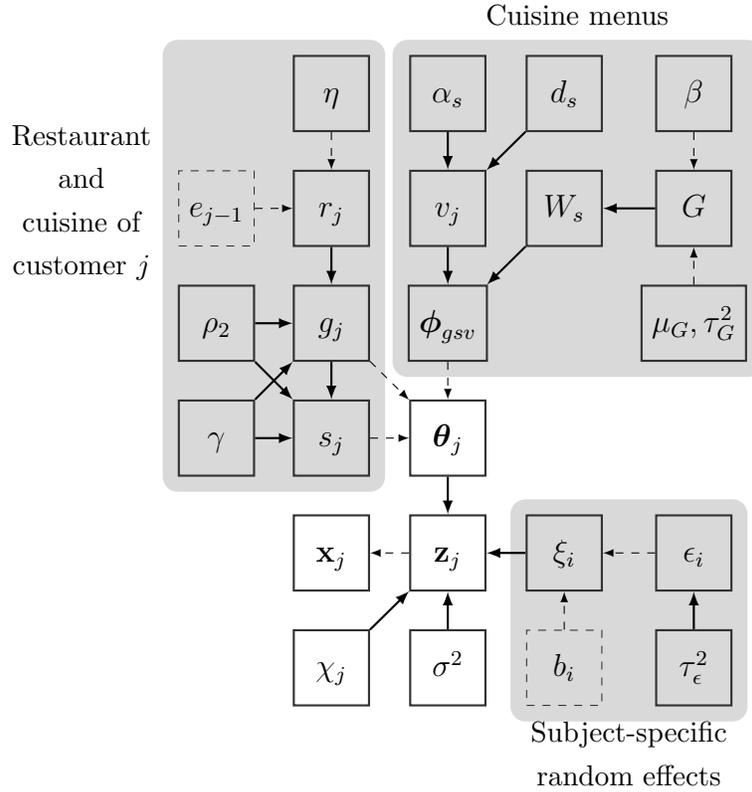
\begin{figure}
\centering


\tikzstyle{stochP}=[rectangle,
                                    thick,
                                    minimum size=1cm,
                                    draw=black!80]
\tikzstyle{modelC}=[rectangle,
                                    dashed,
                                    minimum size=1cm,
                                    draw=black!80]
\tikzstyle{fixedP}=[circle,
                                    thick,
                                    minimum size=1cm,
                                    draw=black!80]
\tikzstyle{textIn}=[rectangle,
                                    dashed,
                                    minimum size=1cm,
                                    draw=white!80]
\tikzstyle{background}=[rectangle,
                                                fill=gray!30,
                                                inner sep=0.2cm,
                                                rounded corners=2mm]

\begin{tikzpicture}[>=latex,text height=1.5ex,text depth=0.25ex]

  \matrix[row sep=0.5cm,column sep=0.5cm] {

 & \node (eta) [stochP]{$\eta$};  &   \node (alphas) [stochP] {$\alpha_s$} ;  & \node (ds) [stochP] {$d_s$}; &  \node (beta) [stochP] {$\beta$}; \\
 \node (ejm1) [modelC]{$e_{j-1}$};   & \node (rj) [stochP]{$r_j$};  &   \node (vj) [stochP] {$v_j$};  &  \node (Ws) [stochP]{$W_s$}; &  \node (G) [stochP]{$G$};  \\
     \node (rho) [stochP]{$\rho_2$};   &  \node (gj) [stochP]{$g_j$};    &  \node (phigsv) [stochP] {$\boldsymbol{\phi}_{gsv}$};  &  &  \node (mugtaug) [stochP] {$\mu_{G},\tau_{G}^{2}$};      \\
    \node (gamma) [stochP]{$\gamma$};   &  \node (sj) [stochP]{$s_j$};  &  \node (thetaj) [stochP]{$\boldsymbol{\theta}_j$};  &  &
        \\
    &  \node (xj)  [stochP] {$\mathbf{x}_j$};   &   \node (zj)  [stochP] {$\mathbf{z}_j$};  & \node (xii)  [stochP] {$\xi_i$}; & \node (epsiloni)  [stochP]{$\epsilon_i$};
        \\
    & \node (chij) [stochP] {$\chi_{j}$}; &  \node (sigma2) [stochP] {$\sigma^{2}$}; & \node (bi) [modelC] {$b_i$}; &  \node (taue) [stochP]{$\tau_{\epsilon}^{2}$}; \\
        };

    \path[->]

        (mugtaug) edge[dashed] (G)
        (beta) edge[dashed] (G)
        (ejm1)   edge[dashed] (rj)
        (G) edge[thick] (Ws)
        (Ws) edge[thick] (phigsv)
        (ds) edge[thick] (vj)
        (alphas) edge[thick] (vj)
        (vj) edge[thick] (phigsv)
        (eta) edge[dashed] (rj)
        (rj) edge[thick] (gj)
        (phigsv) edge[dashed] (thetaj)
        (gamma) edge[thick] (sj)
        (gamma) edge[thick] (gj)
        (gj) edge[thick] (sj)
        (gj) edge[dashed] (thetaj)
        (sj) edge[dashed] (thetaj)

        (thetaj) edge[thick] (zj)
        (sigma2) edge[thick] (zj)
        (zj) edge[dashed] (xj)
        (rho) edge[thick] (gj)
        (rho) edge[thick] (sj)
        (taue) edge[thick] (epsiloni)
        (epsiloni) edge[dashed] (xii)
        (bi) edge[dashed] (xii)
        (xii) edge[thick] (zj)
        (chij) edge[thick] (zj)

	      ;

 \begin{pgfonlayer}{background}
        \node [background,
                    fit=(ejm1)(gamma)(sj)(eta)
                    ,label={[align=center,font=\small]left:{Restaurant \\and \\  cuisine of \\customer $j$}}]{};
              \node [background,
                    fit=(alphas)(phigsv)(mugtaug)(beta)
                    ,label=above:{\small Cuisine menus}] {};
              \node [background,
                    fit=(xii)(bi)(taue)(epsiloni)
                    ,label={[align=center,font=\small, label distance=12pt]below:Subject-specific \\ random effects}] {};

     \end{pgfonlayer}

\end{tikzpicture}
\caption{Directed acyclic graph of the BayesDiff model showing the relationships between the model parameters. Solid rectangles represent the data and model parameters. Dashed rectangles represent predetermined  constants. Solid arrows represent stochastic relationships, and dashed arrows represent deterministic relationships. }
\label{DAG}
\end{figure}


\newpage

\bibliographystyle{agsm}

\bibliography{BayesDiff}
\end{document}